\documentclass[preprint,aps,tightenlines,floatfix,showpacs]{revtex4}

\usepackage{graphics}
\usepackage{bm}
\usepackage{amsmath}
\usepackage{amssymb}
\begin{document}

\title
{Comparison of optical model results from a microscopic Schr\"odinger 
approach to nucleon-nucleus elastic scattering with those from a global
Dirac phenomenology.}
\author{P. K. Deb}
\email{pdeb@mps.ohio-state.edu}
\author{B. C. Clark}
\email{bcc@mps.ohio-state.edu}
\affiliation
{Department of Physics, The Ohio State University, Columbus, OH 43210,
U.S.A.}
\author{S. Hama}
\email{sn-hama@hue.ac.jp}
\affiliation
{Hiroshima University of Economics, Hiroshima 731-0192, Japan}
\author{K. Amos}
\email{amos@physics.unimelb.edu.au}
\author{S. Karataglidis}
\email{kara@physics.unimelb.edu.au}
\affiliation
{School of Physics, The University of Melbourne, Victoria 3010, 
Australia}
\author{E. D. Cooper}
\email{tim.cooper@ucfv.ca}
\affiliation
{University College of the Fraser Valley, Abbotsford, BC 72S 7M8 Canada}

\date{\today}

\begin{abstract}
Comparisons are made between results of  calculations  for  intermediate energy 
nucleon-nucleus scattering for ${}^{12}$C, ${}^{16}$O, ${}^{40}$Ca, ${}^{90}$Zr,
and   ${}^{208}$Pb,  using  optical   potentials  obtained  from  global  Dirac 
phenomenology and from a microscopic Schr\"odinger model.    Differential cross 
sections  and  spin  observables for  scattering from the set of five nuclei at 
65~MeV  and  200~MeV  have  been  studied to assess the relative merits of each 
approach.     Total reaction cross sections from proton-nucleus and total cross
sections from  neutron-nucleus scattering have been evaluated and compared with 
data for those five targets in the energy range 20 MeV to 800 MeV.  The methods
of  analyses  give  results  that  compare well with experimental data in those 
energy regimes for which the procedures are suited.
\end{abstract}
\maketitle

\section{Introduction}

   Traditionally the physics of the interaction of a nucleon with a nucleus has 
been represented by an optical potential. Once that potential is specified, and
by its use also the scattering matrix, all observables may be calculated.   For 
intermediate  energies  there  have been a number of methods of calculating the 
optical potential,  the  most  successful  of  which have been the global Dirac 
phenomenology (DP)~\cite{Ra92} and a microscopic Schr\"odinger model~\cite{Am00}
in coordinate space identified in the literature as the $g$-folding method. 

The global DP approach is the outgrowth of the seminal work by Walecka and  his 
group~\cite{Se86} with a second order reduction of the Dirac  equation  leading
to Schr\"odinger equivalent equations.  Potentials in those equations have been
found   with  which  good  fits  to  elastic  nucleon-nucleus  scattering  data 
result~\cite{Ra92} and more recently,  using  a  global DP, very good fits have
been  found  for nucleon  energies ranging from 20~MeV to 1040~MeV~\cite{Co93}.
Such  potentials  are utilitarian  for use in analyses of other data.    In the 
$g$-folding approach,  optical  potentials  are  obtained  by first deriving an 
effective $NN$ interaction  in-medium  from  a realistic  free  nucleon-nucleon 
($NN$) potential.  That effective interaction is then folded  with  a  suitable
representation of the density matrix of the target nucleus.

Both approaches have had success. Not only have they given good predictions  of
differential  cross sections  but  also  they have reproduced spin and integral 
observables~\cite{Am00,Ko89,De01,Co93}.  A specific example of the success with
the  $g$-folding   method  has  been  its  use  in  analyzing 200~MeV data from 
${}^{208}$Pb.    Excellent  results  were  obtained  with  structure given by a 
Skyrme-Hartree-Fock  (SHF)  model~\cite{Br00}  for  the  ground  state  that is
consistent   with   the    Friedman-Pandaharipande    neutron    equation    of 
state~\cite{Fr81}.  The global DP approach,  based  on the relativistic impulse
approximation (RIA), fits the  data well when proton  and  neutron  densities, 
their root-mean-square radii, and the neutron skin thickness, all are credible.
Targets of $^{40}$Ca, $^{48}$Ca and $^{208}$Pb were considered~\cite{Cl03}.

However, the two models fundamentally are different.     The global DP model is
dependent on the fitting of data to determine the  parameters  in  the  assumed
potentials.  Conversely,  the microscopic Schr\"odinger model is dependent upon
obtaining  realistic  effective  $NN$  interactions and upon the quality of the 
model chosen for the structure of the ground state of the target.     A complex 
non-local optical  potential results. Of course one can find a local equivalent
to  this  but  there  is  no  {\em  a  priori}   reason  to   assume  that  the 
(non-relativistic) potential deduced  from  the Dirac reduction resembles,   in 
any way, that equivalent local potential. 
 
Yet, to date, only one real comparison~\cite{Ka01} has been made simultaneously 
of  the  results  from  the  Dirac  and  the $g$-folding approaches for nucleon 
projectiles, and that for  the  ${}^{208}$Pb nucleus and  100, 200, and 300~MeV
nucleons.    Also,  an  earlier  comparison  of the Dirac and the then  favored 
phenomenological  Schr\"odinger  models  revealed   problems  inherent  in  the 
latter~\cite{Ko89}.  The phenomenological Schr\"odinger model usually assumes a 
simple local Woods-Saxon form of the optical potential.  Such an approach often
fails  to reproduce  spin  observables  well  enough due to assumptions made in 
defining the spin-orbit potential.    However,  a more complex phenomenological 
(local) Schr\"odinger model has been developed recently~\cite{Ko03},     with a 
global parameter specification, that does replicate very well a large  data set 
for energies 0 to 200~MeV.

     The past successes of the two model prescriptions suggest that both may be 
preset in a way that allows a fair comparison between them,   particularly when 
the results are predictions in that no  {\it a posteriori}  adjustments to  any 
facet of the base model details are considered.    With that in mind, the prime 
purpose of this paper is to  compare in detail results obtained from the global 
DP model with those from the $g$-folding model for proton scattering.  First we 
report on results of predictions of angular observables for elastic  scattering
of  65~MeV  and of  200~MeV  protons  from ${}^{12}$C, ${}^{16}$O, ${}^{40}$Ca, 
${}^{90}$Zr, and ${}^{208}$Pb.   For all cases, the results of the calculations
made were plotted before any data was added.          We also consider integral 
observables for both proton and neutron scattering for energies  to  800~MeV to
evaluate the isospin dependence of the potentials.  
 
  In this paper,  the formalism of the microscopic $g$-folding and that of the 
global DP approach  are  discussed  briefly  in  Sect.~II  and the results are 
presented in Sect.~III.  In the latter there are two subsections. In the first 
of those,  the  results of calculations at 65~MeV and 200~MeV of  differential
cross sections and of spin observables,  $A_y$ and $Q$, are compared with data. 
In the second,  predictions  of reaction and total reaction cross sections made 
using both methods are compared  with data for the five nuclei selected to span 
the mass range. Conclusions that may be drawn are given thereafter in Sect.~IV.

\section{Models of the nucleon-nucleus optical potential}

 As detailed presentations of the two model prescriptions of interest have been 
published~\cite{Ra92,Co93,Am00}, only salient features are given herein. 

\subsection{The microscopic $g$-folding model}

    To form the $g$-folding model optical potential it is assumed that pairwise 
interactions between the projectile nucleon and each bound nucleon as prescribed
by a large basis model of the target suffice.         Thus,  in the $g$-folding 
approach~\cite{Am00}, no phenomenological form for the potential is assumed nor 
is that potential derived from any other phenomenological potential.    Instead
the method begins with the $g$-matrices of a realistic $NN$ potential;    those 
$g$-matrices being solutions  of  the  Brueckner-Bethe-Goldstone  equations  in 
infinite nuclear matter,
\begin{align}
g_{L L'}^{J S T} (p', p; k, K, k_{F})  & =  V_{L L'}^{J S T}(p, p') 
\,+\,\frac{2}{\pi} \sum_{l} \int^{\alpha}_0 V_{L l}^{J S T} (p', q)
\nonumber\\
&\times \left\{  \frac{\bar{Q} (q,K,k_{f})}{\bar{E}(k,K,k_{f}) - \bar{E}
(q,K,k_{f}) + i\epsilon} \right\}
\,g_{l L'}^{(J S T)}(q,p;k,K,k_{F})\,q^{2}\;dq\ , 
\end{align}
in  which  $\bar{Q}(q,K,k_{f})$  is  an  angle-averaged  Pauli operator with an 
average  center-of-mass momentum $K$.    The energies,  $\overline E$,  in  the 
denominator are single particle values within  an  averaged  field~\cite{Am00}.
Such equations are solved for 32 $NN$ channels, for Fermi momenta ($k_F$) up to
densities 1.5 times the nuclear density, and for many energies ($k$ found from
$E$ with relativistic kinematics)
to  300~MeV  using  the  Bonn-B  $NN$  potential~\cite{Ma87}  as  the  starting 
interaction for the calculations.

 To form optical potentials in coordinate space then necessitates a  mapping of 
those  $g$-matrices  onto  Bessel  transforms  of  a  complex  $NN$   effective 
interaction.      That  has  been  made  with  particular emphasis placed  upon 
reproduction of the on-shell values of the $g$-matrices.          The effective 
interaction is the mix of sets of  Yukawa  function  form factors  for central, 
two-nucleon spin-orbit, and tensor operators that  can  be  used   in  the code 
DWBA98~\cite{Ra99}.  The density of the input structure for the  target  ground 
state is used  to  select  what  values  of  the  complex and  energy dependent 
strengths of the effective interaction are used in the folding process.    When 
folded with an appropriate ground state density of the target, the  microscopic 
optical potential is obtained naturally,   incorporating   Pauli  blocking  and 
density dependences. The (coordinate space) potential  contains both direct and 
exchange parts; the latter arising from   antisymmetrization  of the projectile
and bound state nucleon wave functions.    Consequently the  potential is fully 
nonlocal. As formulated~\cite{Am00}, there are no parameters  in the  model  to 
adjust {\em a posteriori}.    All  results  then  are  obtained  from  a single 
(predictive) calculation.      For energies at which this theory is applicable, 
much success has been achieved   predicting the observables from proton-nucleus 
scattering for a number of nuclei~\cite{Am00}. 

Formally, and ignoring  complications of spin interactions and expectations, we
seek solutions of the Schr\"odinger equations of the form
\begin{equation}
\left[ \frac{\hbar^2}{2\mu} \nabla^2 - V_{c}(r) + E \right]\,
\Psi({\bf r}) = \int U({\bf r}, {\bf r'}; E)\,\Psi({\bf r'})\; d{\bf r'} \ ,
\label{Schr1}
\end{equation}
where  ${\bf r},  {\bf r'}$  are  relative  $NA$  coordinates, $V_{c}(r)$ is  a 
Coulomb interaction and $U({\bf r},{\bf r'})$ is the optical potential.      In 
practice, such potentials, or their underlying multipoles,  are  not  evaluated 
specifically when DWBA98  is  used.      However,  it  is  useful  to  consider
$U({\bf r},{\bf r'})$ and  how  it  may  be  formed  to  note  how  the  target 
structure enters the process.             Consider the overlaps which  lead  to 
$U({\bf r},{\bf r'})$, i.e.
\begin{equation}
U_{pA}(0,1) = 
\left( \Psi_{gs}(1,2,\cdots,A) \left| \sum_{n=1}^A g_{eff}(0n) \right|
\Psi_{gs}(1,2,\cdots,A) \right) 
= \left( \Psi_{gs} \left| A g_{eff}(01) \right| \Psi_{gs} \right) \ ,
\label{Overlap}
\end{equation}
where `0' denotes the projectile coordinates, and as all nucleons in the target
are equivalent, we have chosen a specific entry (`1').  By so doing we can make
use of a cofactor expansion of the nuclear many-body (ground) state,
\begin{equation}
\left| \Psi(1,2,\cdots,A) \rangle \right.\,=\,\frac{1}{\sqrt{A}} 
\sum_{\alpha m} \left| \varphi_{\alpha m}(1) \rangle \right.\, 
{\underline{a_{\alpha m} \left| \Psi(1,2,\cdots,A) \rangle \right.}}
\ .
\label{state}
\end{equation}
Here  $\alpha$   specifies  the  quantum  number set \{$n,l,j, \zeta$\}  where 
$\zeta$ is the isospin projection.     Then,  as  the  underlined   factor  in 
Eq.~(\ref{state}) is independent of coordinate `1',
\begin{equation}
U_{pA}(0,1) = \sum_{\alpha m \alpha' m' } \langle \Psi |
a_{\alpha' m'}^{\dag} a_{\alpha m} | \Psi \rangle\, 
\left( \varphi_{\alpha' m'}(1) \right| g_{eff}(10) \{ \left| 
\varphi_{\alpha m}(1) \right) - \left| \varphi_{\alpha m}(0) \right) 
\} \ ,
\end{equation}
when  the  required antisymmetry with projectile and struck nucleon  is taken 
into account.

   The many-body reduced matrix  elements of the particle-hole operator pairs 
are the OBDME~\cite{Am00}.  Usually for elastic scattering they are just the 
nucleon shell occupancies of the target, $\eta_\alpha$. Thus such occupancies, 
the  single  nucleon  bound  (SP)  states,  and  the  effective  interactions 
$g_{eff}(01)$, define the optical potential in this approach, as
\begin{align}
U({\bf r}_{1},{\bf r}_{2} ; E) = \sum_{\alpha m}  \eta_\alpha 
\,\Biggl[ \delta ({\bf r}_{1} - {\bf r}_{2})\, 
&\left.\int \varphi_{\alpha m}^{*}({\bf s})
\,U^{D}(R_{1s},E)\,\varphi_{\alpha m}({\bf s})\;d{\bf s} \right.
\nonumber\\
&+\,\varphi_{\alpha m}^{*}({\bf r}_{1})\, 
U^{Ex}(R_{12},E)\,\varphi_{\alpha m}({\bf r}_{2}) {\bigg ]} \ ,
\end{align}
where $R_{12} = | {\bf r}_{1} - {\bf r}_{2} |$, and $U^{D}$ and $U^{Ex}$  are 
appropriate combinations of components of the effective interaction for   the 
direct    and    exchange    contributions    to    the    optical  potential 
respectively~\cite{Am00}.

\subsection{Global DP optical model potentials}

A second-order reduction of the Dirac equation for nucleon-nucleus leads to a 
Schr\"odinger-equivalent  equation  that  has  physically correct, effective,
central and spin-orbit potentials.   Symmetry  allows  one  to  have  Lorentz 
scalar, S, Lorentz vector, V,  and tensor, T potentials which,  collectively,  
are termed SVT potentials.   However, the tensor force can always be replaced 
in a phase equivalent way by an effective relativistic potential of just scalar 
and vector parts, and as our approach is phenomenological,  we  consider  the  
global  DP  using scalar-vector (SV) potentials, together  with  the  Coulomb 
potential.  Solutions  of those  Schr\"odinger-equivalent equations have given  
$S$-matrices with which observables have been predicted very well~\cite{Cl85}.  
In  the  SV, the real vector potential is large and repulsive  while the real 
scalar potential is somewhat larger and  attractive.     The imaginary vector 
potential  is  attractive  and  the imaginary scalar is repulsive.  The Dirac 
equation so structured is suitable for simultaneous analyses of proton-nucleus
and  neutron-nucleus scattering data for incident energies up to several GeV.

The  scalar-vector (SV) model of global DP that has been fit to   the elastic 
proton-nucleus scattering observables~\cite{Co93} takes the form
\begin{align}
U(r, E, A)\,=\,V^v(E,A)&\,f^v(r,E,A)\, +\, V^s(E,A)\,f^s(r,E,A)
\nonumber \\
& +\,iW^v(E,A)\,g^v(r,E,A)\, +\, iW^s(E,A)\,g^s(r,E,A)\ ,
\label{eqn1}\,
\end{align}
The  superscripts  $v$  and  $s$  refer  to volume  and  surface peaked terms 
respectively. The ``COSH'' form~\cite{Co93} is used in this paper, {\em viz}.
\begin{eqnarray}
f^v  &=& \frac{\left\{ \text{cosh} \left[ R(E,A)/a(E,A)\right]-1
\right\}} {\left\{ \text{cosh} \left[ R(E,A)/a(E,A)\right]\, +\,
\text{cosh} \left[ r/a(E,A) \right] - 2 \right\} } ,
\nonumber\\
&&\nonumber\\
f^s &=& \frac{\left\{ \text{cosh} \left[ R(E,A)/a(E,A)\right]\,-\,1
\right\} \left\{ \text{cosh} \left[ r/a(E,A)\right]\, -\, 1\right\}}
{\left\{ \text{cosh} \left[ R(E,A)/a(E,A)\right]\, +\, \text{cosh} \left[
 r/a(E,A) \right]\, -\, 2 \right\}^2 } .
\end{eqnarray}
Similar forms have been taken for $g^v$ and $g^s$. Thus the optical potential 
consists of scalar and vector terms each having real and imaginary parts.  In 
the global analyses~\cite{Co93} the eight strength parameters varied with the 
proton center of mass energy $E$ (in MeV) and  with the  atomic  mass  of the 
target $A$. The results can be represented by polynomials,
\begin{equation}
V^v(E,A) = v_0 + \sum_{m=1}^{4}v_mx^m + \sum_{n=1}^{3}v_{n+4}y^n
+ v_8xy + v_9x^2y + v_{10}xy^2,
\end{equation}
where  $x = 1000/E$  and $y = A/(A+20)$.   We have used the same form for all 
other potentials $V^s(E,A)$, $W^v(E,A)$, and  $W^s(E,A)$.  But the scalar and 
vector  potentials,  and  their  real  and  imaginary  parts,  all  may  have 
different geometry parameter values,  the energy and mass dependence of which 
can be expressed by
\begin{eqnarray}
R &=& A^{1/3} \left[ r_0 + \sum_{m=1}^{4}r_mx^m +
\sum_{n=1}^{3}r_{n+4}y^n + r_8xy + r_9x^2y + r_{10}xy^2 \right],
\nonumber\\
a &=& a_0 + \sum_{m=1}^{4}a_mx^m + \sum_{n=1}^{3}a_{n+4}y^n + a_8xy
+ a_9x^2y + a_{10}xy^2.
\end{eqnarray}
There  are  many  parameters  in  this specification which  have been defined 
suitably by using a large data set in a search process.      In the global DP 
calculations  we  use  the  recoil  correction  given  by   Cooper        and 
Jennings~\cite{Co88}. 

\subsection{Phase shifts, $S$-matrices, and observables}

Irrespective of how the $NA$ optical potential is specified, the objective of 
its  use  is  to  define  the  scattering  ($S$)-matrix,  or equivalently the 
(complex)  phase shifts  $\delta^{\pm}_l(k)$.     With  superscripts  ($\pm$) 
identifying $j= l\pm 1/2$, these relate by
\begin{equation}
S^{\pm}_l(k) = e^{2i\delta_l^{\pm}(k)} = \eta^{\pm}_l(k)
e^{2i\Re\left[ \delta^{\pm}_l(k) \right]}
\hspace*{0.5cm}{\rm where}\hspace*{0.5cm}
\eta^{\pm}_l(k) = \left| S^{\pm}_l(k) \right| =
e^{-2\Im\left[ \delta^{\pm}_l(k) \right]} \; .
\end{equation}
With  $E \propto k^2$,  integral observables for neutrons,  namely the total 
elastic, total reaction (absorption), and total cross sections then are given 
by
\begin{eqnarray*}
\sigma_{\text{el}}(E) & = & \frac{\pi}{k^2} \sum^{\infty}_{l = 0} \left\{
\left( l + 1 \right) \left| S^+_l(k) - 1 \right|^2 + l \left|
S^-_l(k) - 1 \right|^2 \right\} \; , \\
\sigma_{\text{R}}(E) & = & \frac{\pi}{k^2} \sum^{\infty}_{l = 0} \left\{
\left( l + 1 \right) \left[ 1 - \eta^+_l(k)^2 \right] + l \left[ 1 -
\eta^-_l(k)^2 \right] \right\} \; ,
\end{eqnarray*}
and
\begin{eqnarray}
&&\sigma_{\text{TOT}}(E)  =  \sigma_{\text{el}}(E) +
\sigma_{\text{R}}(E)\nonumber \\
&&\hspace*{0.5cm} = \frac{2\pi}{k^2} \sum^{\infty}_{l = 0} 
\left\{ \left( l + 1
\right) \left[ 1 - \eta^+_l(k)\cos\left( 2\Re\left[ \delta^+_l(k)
\right] \right) \right] + l \left[ 1 - \eta^-_l(k) \cos\left(
2\Re\left[ \delta^-_l(k) \right] \right) \right] \right\}, \ \ 
\end{eqnarray}
respectively.      Note that the Coulomb interaction means that only total 
reaction  cross sections  (formed by a modification of the above)  can  be 
measured from proton scattering.

To evaluate angular observables,   and specifically the differential cross 
sections and analyzing powers,  one  need  form  the scattering amplitudes 
which, for nucleon elastic scattering are then  $2 \times 2$  matrices  in 
the nucleon spin space having the form,
\begin{equation}
f(\theta) = A(\theta) + i B(\theta) \bm{\sigma} \cdot \hat{\bm{n}}\; ,
\label{spinamp}
\end{equation}
where, in terms of the $S$-matrix elements~\cite{La90}, 
\begin{eqnarray}
A(\theta) & = & \frac{1}{2ik} \sum_{l=0} \left\{ \left( l + 1 \right)
\left[ S^+_l(k) - 1 \right] + l \left[ S^-_l(k) - 1 \right] \right\}
P_l(\theta) \; , \nonumber \\
B(\theta) & = & -\frac{1}{2ik} \sum_{l=1} \left[ S^+_l(k) - S^-_l(k)
\right] P_l^1(\theta) \; .
\end{eqnarray}
From these (complex) amplitudes, and with the quantization axis normal to
the scattering plane, the (elastic scattering) differential cross 
section, analyzing power, and spin rotation are defined by
\begin{eqnarray}
\frac{d\sigma}{d\Omega} &=& \left| A(\theta) \right|^2 + \left|
B(\theta) \right|^2\nonumber\\
A_y(\theta) &=& \frac{2\Im\left[A^*(\theta)B(\theta)\right]}
{d\sigma/d\Omega}\nonumber\\
Q(\theta) &=& \frac{2\Re\left[A^*(\theta)B(\theta)\right]}
{d\sigma/d\Omega}\; .
\end{eqnarray}

The $g$-folding model results have been obtained using the DWBA98 code~\cite{Ra99}
which evaluates scattering amplitudes using a helicity formalism~\cite{Am00}.
Those scattering amplitudes ${\hat A}(\theta), {\hat B}(\theta)$ relate to 
$A(\theta), B(\theta)$ by a unitary transformation involving the rotation matrices 
$\{{\bf d}^{\frac{1}{2}}(\theta)\}$ under which the cross section  and analyzing 
power are invariant, in that
\begin{eqnarray}
\frac{d\sigma}{d\Omega} &=& \left| A(\theta) \right|^2 + \left|
B(\theta) \right|^2 \equiv \left| {\hat A}(\theta) \right|^2 + \left|
{\hat B}(\theta) \right|^2\nonumber\\
A_y(\theta) &=& \frac{2\Im\left[{\hat A}^*(\theta){\hat B}(\theta)\right]}
{d\sigma/d\Omega}
\end{eqnarray}
but the spin rotation becomes
\begin{equation}
Q(\theta) = \frac{1}{d\sigma/d\Omega}
\left[
\left(\left|{\hat A}(\theta)\right|^2 - \left|{\hat B}(\theta)\right|^2\right) 
\sin(\theta)\ 
+\  
\left(2\Re\left[{\hat A}^*(\theta) {\hat B}(\theta)\right] \right) \cos(\theta)
\right]\ .
\end{equation}

\section{Results and data comparisons}

  We have chosen five nuclei for study.       They are ${}^{12}$C,  ${}^{16}$O,
${}^{40}$Ca, ${}^{90}$Zr, and ${}^{208}$Pb. Not only do they span a large range
of target mass and ground state isospin,   but also some have been described by 
mean fields derived from large space structure studies which suit comparison of
scattering models.

     All microscopic model results we show have been evaluated using the DWBA98 
program~\cite{Ra99},   input  to  which  are  the density dependent and complex 
effective $NN$ interactions described earlier.    Other input to DWBA98 are the 
ground state occupancies (OBDME) and the associated SP functions; both of which
are defined from nuclear structure calculations.    The structures we have used
for each nucleus in the $g$-folding calculations are
\begin{itemize}
\item ${}^{12}$C:
    For this,    the lightest mass nucleus we consider,  a  no-core shell model 
calculation~\cite{Ka95}  in  the  full  $(0+2)\hbar \omega$  space  defined its 
ground state.  The WBT interaction of Warburton and Brown~\cite{Wa92} was used. 
With exception of the  $0^+_2$ state,  this shell model calculation gave a very 
good spectrum to 20 MeV excitation.
\item ${}^{16}$O:
    For this nucleus,  we have used the structure as determined by  Haxton and 
Johnson~\cite{Ha90}. That was a $(0+2+4)\hbar \omega$ calculation made using a
hybrid interaction based upon  the  $0p$-shell  Cohen  and  Kurath~\cite{Co65} 
potentials.   The  (positive parity)  spectrum  of ${}^{16}$O  to about 10 MeV 
excitation is quite well reproduced.
\item ${}^{40}$Ca:
    The  ground  state  structure  for  this  nucleus  was  taken  from recent 
studies~\cite{Br00,Ka02} made using the SHF(SKX) model of structure. That model
lead to better fits to the 200 MeV data (which we consider again herein)  than 
either the simple $0\hbar \omega$ shell model or SHF results made  with  other
interactions.  
\item ${}^{90}$Zr:
                          For this nucleus, a shell model space defined in the 
$1p_{\frac{3}{2}}, 0f_{\frac{5}{2}}, 1p_{\frac{1}{2}}, 0g_{\frac{9}{2}}$ orbits
was used with the NIS interaction~\cite{Ji89}.  That  interaction was optimized
to describe nuclei in the Ni$\to$Sn mass region. 
\item ${}^{208}$Pb:
This ground state of this nucleus also is described by a SHF calculation.   In 
this case, using the SKM* interaction~\cite{Br00},  wave  functions and  OBDME 
(shell occupancies) result with which the  40,  65, and 200 MeV proton elastic 
differential cross sections as well as the  electron  scattering  form  factor
have been extremely well reproduced~\cite{Ka02}.
\end{itemize}  
As noted before~\cite{Am00}, with all details preset, the $g$-folding approach
is predictive.  So only one calculation of an optical potential and of its use
to give cross sections and spin observables is made. How good a reproduction of
data results then is a measure of the formulation and/or of the specifications 
used.

The global Dirac approach~\cite{Co93} is different in philosophy to the above.  
First  it  is  phenomenological  and  the functional forms of parameter values 
were obtained by fitting a large body of data, including most if not all of the 
data considered herein.     From  those  studies, two parameter sets, which we 
identify   as   the   $EDAI$   ($E$-dependent  $A$-independent)   and   $EDAD$ 
($E$-dependent $A$-dependent), specify the Dirac potentials we shall use. From 
the general form of the model given in Eq.~(\ref{eqn1}),   there   are   eight
potentials which are functions of only energy and radius. The $EDAD$ fits test
the  sensitivity  of  the  computation  to the input optical model potentials;
though there are three different  $EDAD$  fits~\cite{Co93}  that  give equally
high-quality agreement with the data.

In all figures to follow, results found using the $g$-folding, the $EDAI$, and 
the $EDAD3$ models will be displayed by the solid, long-dashed, and dot-dashed 
curves respectively.

\subsection{Cross sections and spin observables: angular variations.}

\begin{figure}
\scalebox{0.7}{\includegraphics{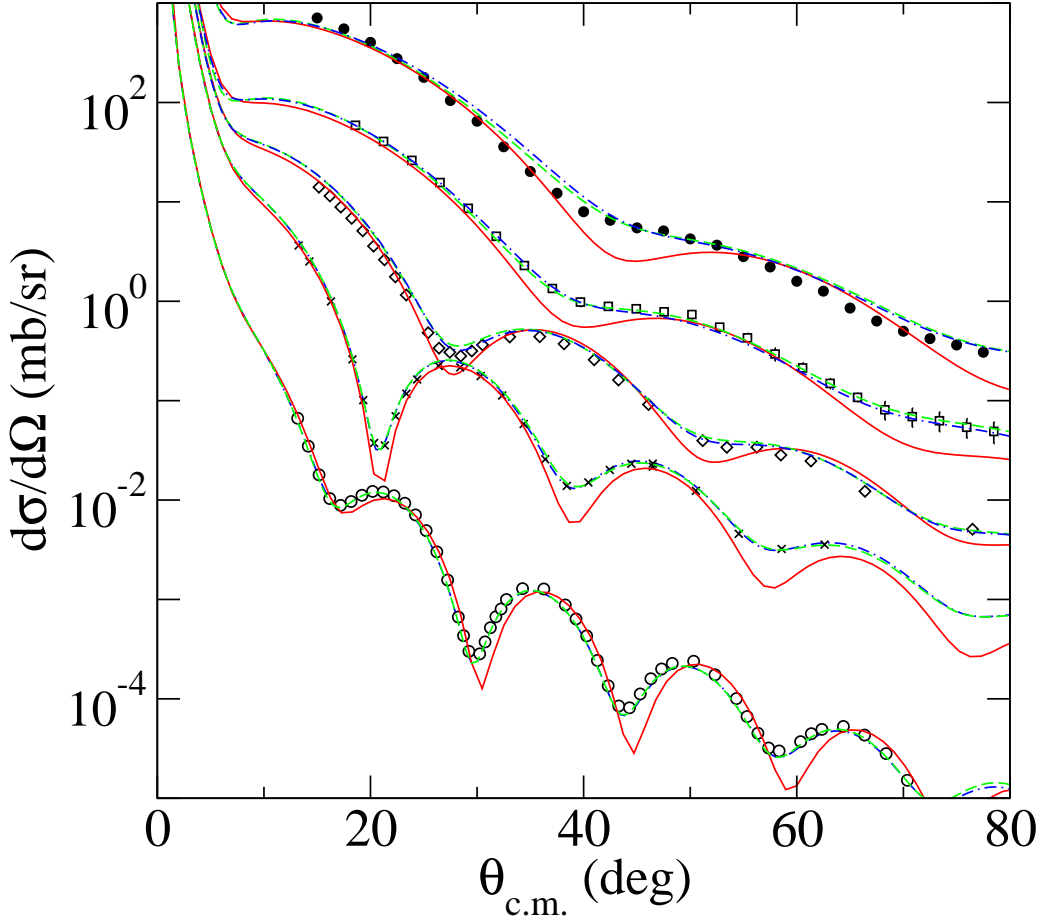}}
\caption{\label{xsec65}(Colour online)
The differential cross sections from the scattering of 65 MeV protons from 
five nuclei. Data~\cite{No81,Sa82,Ka85} are compared with the results of 
different theoretical model calculations.  Details are given in the text.}
\end{figure}

\begin{figure}
\scalebox{0.7}{\includegraphics{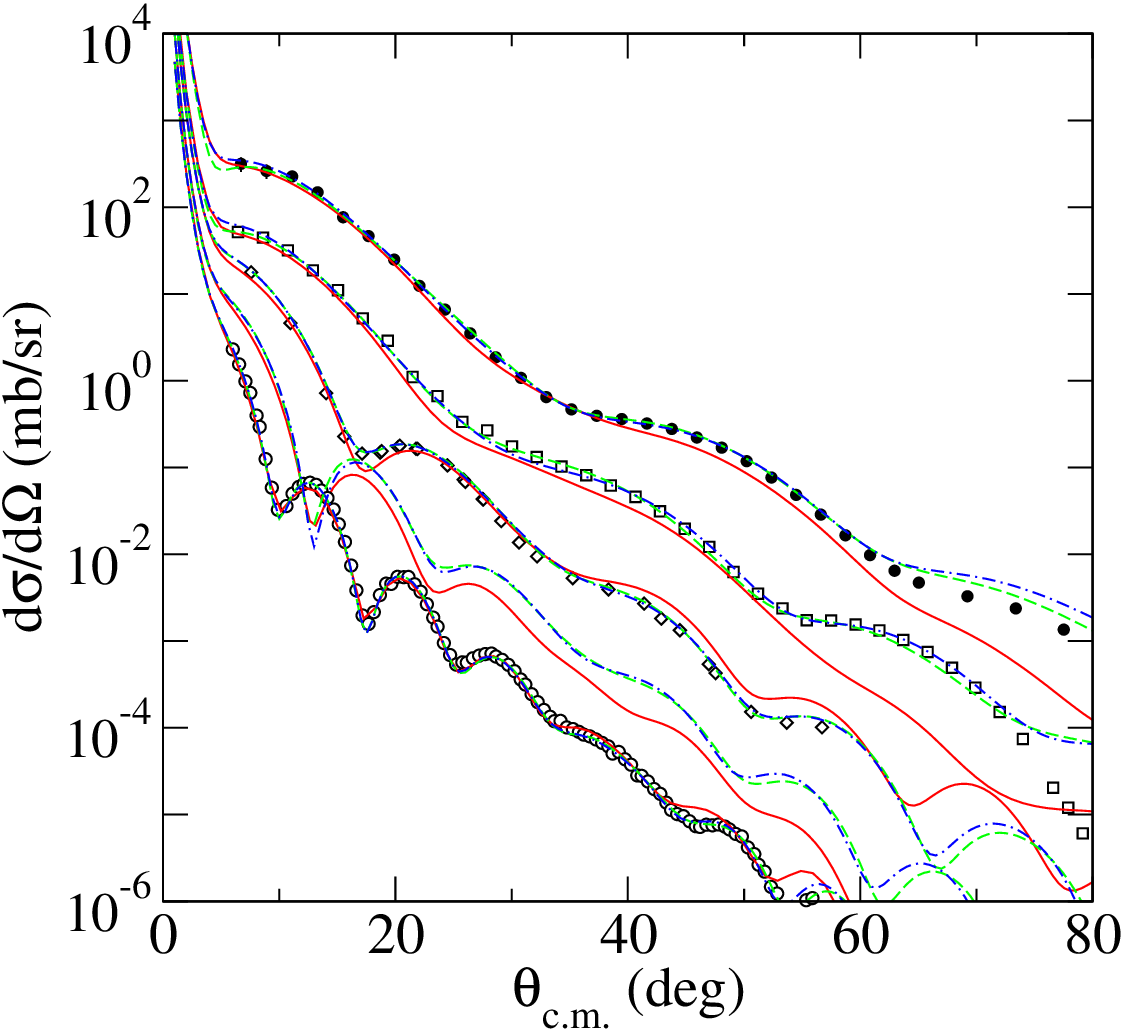}}
\caption{\label{xsec200}(Colour online)
The differential cross sections from the scattering of 200 MeV protons from 
five nuclei. Data~\cite{Co82,Si93,Hu81} are compared with the results 
of different theoretical model calculations.  Details are given in the text.}
\end{figure}

    The differential cross sections for 65~MeV and 200~MeV protons elastically
scattered   from   ${}^{12}$C,   ${}^{16}$O,   ${}^{40}$Ca,  ${}^{90}$Zr,  and 
${}^{208}$Pb are shown in Figs.~\ref{xsec65}  and  \ref{xsec200} respectively.
The calculated results are compared with data in nine of the ten cases.    The 
data are displayed by different symbols: filled circles for ${}^{12}$C,   open 
squares for ${}^{16}$O, open diamonds for ${}^{40}$Ca, crosses for ${}^{90}$Zr,
and open circles for ${}^{208}$Pb.   With  all  results  and  data  scaled for 
clarity, from the top down in Figs.~\ref{xsec65} are the results and data  for 
${}^{12}$C (1),   ${}^{16}$O ($10^{-1}$),   ${}^{40}$Ca ($10^{-2}$), 
${}^{90}$Zr ($10^{-3}$), and ${}^{208}$Pb ($10^{-5}$). The scales used are 
given in brackets following the target symbol.  

 Clearly both global DP results fit the data very well as the studies
to determine the global set were designed to do.   The  $g$-folding results, by
comparison, do not do as well.   But recall that each  $g$-folding result is  a
prediction built upon a given microscopic model of the target structure  and an
effective interaction determined from infinite matter  solutions  of  the  $NN$
$g$-matrices.  As such, these results also are very good, predicting reasonable
structures and essentially the correct magnitudes for all cases. The structures
of  the  $g$-folding results are sharper than those of the data suggesting that 
the absorptive character of the $g$-folding optical potentials at 65~MeV may be
a little weak.

   Such  is  far  less the case with the 200 MeV results which are displayed in
Fig.~\ref{xsec200}.   In  this  figure the notation is as for Fig.~\ref{xsec65}
with  now  the scale for the ${}^{208}$Pb results (lowest in the diagram) being
$10^{-4}$.
Again both of the global DP cross sections give very good fits to the 
data while now the  $g$-folding  predictions  are  good representations of that
data though with some noticeable disparities  at  the larger scattering angles;
where the magnitudes of the cross sections  are  less than  0.1~mb/sr  and  the 
momentum transfer values are in excess of $\sim 2.5$ fm$^{-1}$.        For such
conditions,  the  first  order theory upon which the $g$-folding model is built 
may be too limited~\cite{Am00}.

\begin{figure}
\scalebox{0.7}{\includegraphics{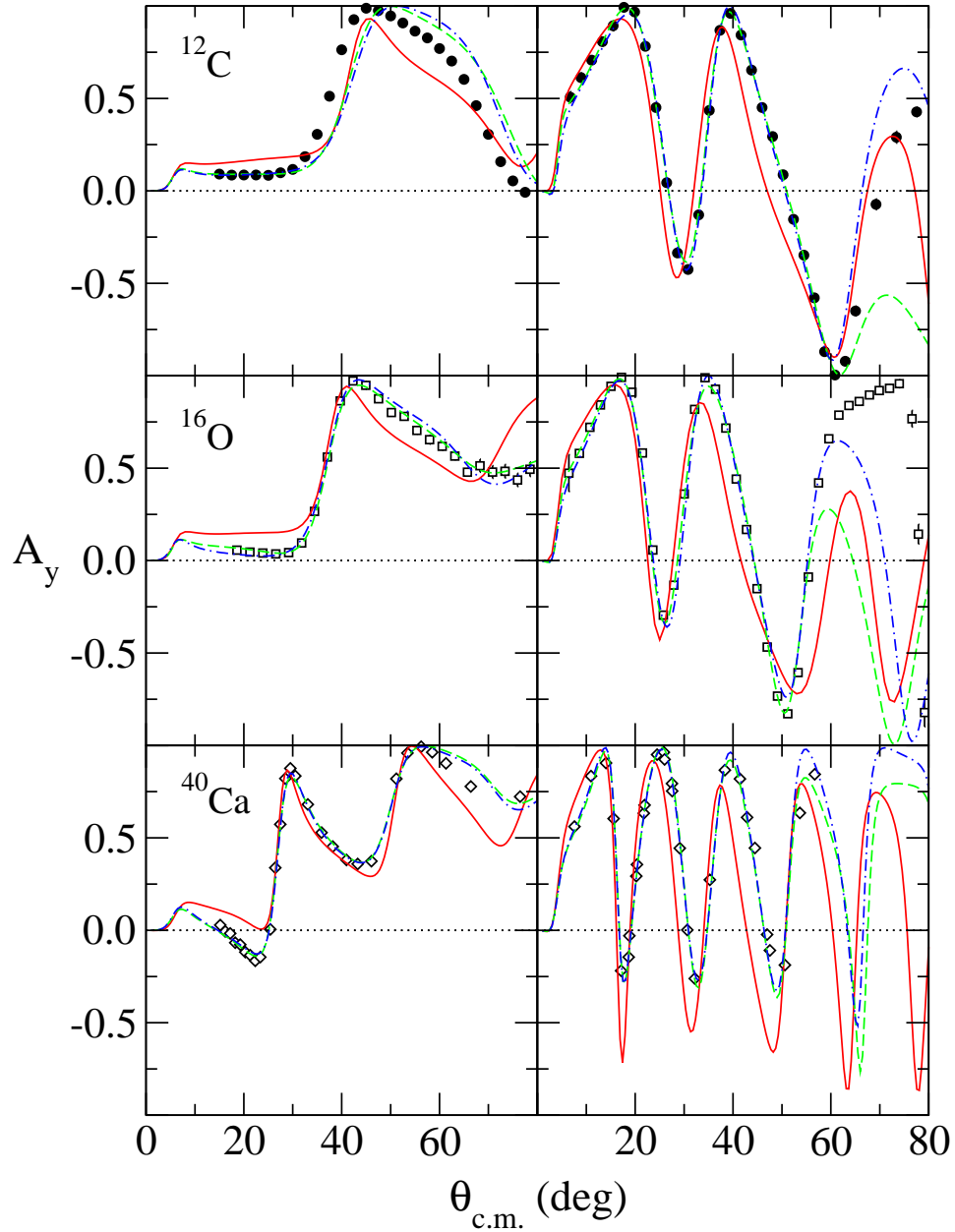}}
\caption{\label{CCa-Ay}(Colour online)
Analyzing powers from 65 (left) and 200 (right) MeV proton elastic
scattering from ${}^{12}$C (top), ${}^{16}$O (middle), and ${}^{40}$Ca (bottom).
Details are given in the text.}
\end{figure}

\begin{figure}
\scalebox{0.7}{\includegraphics{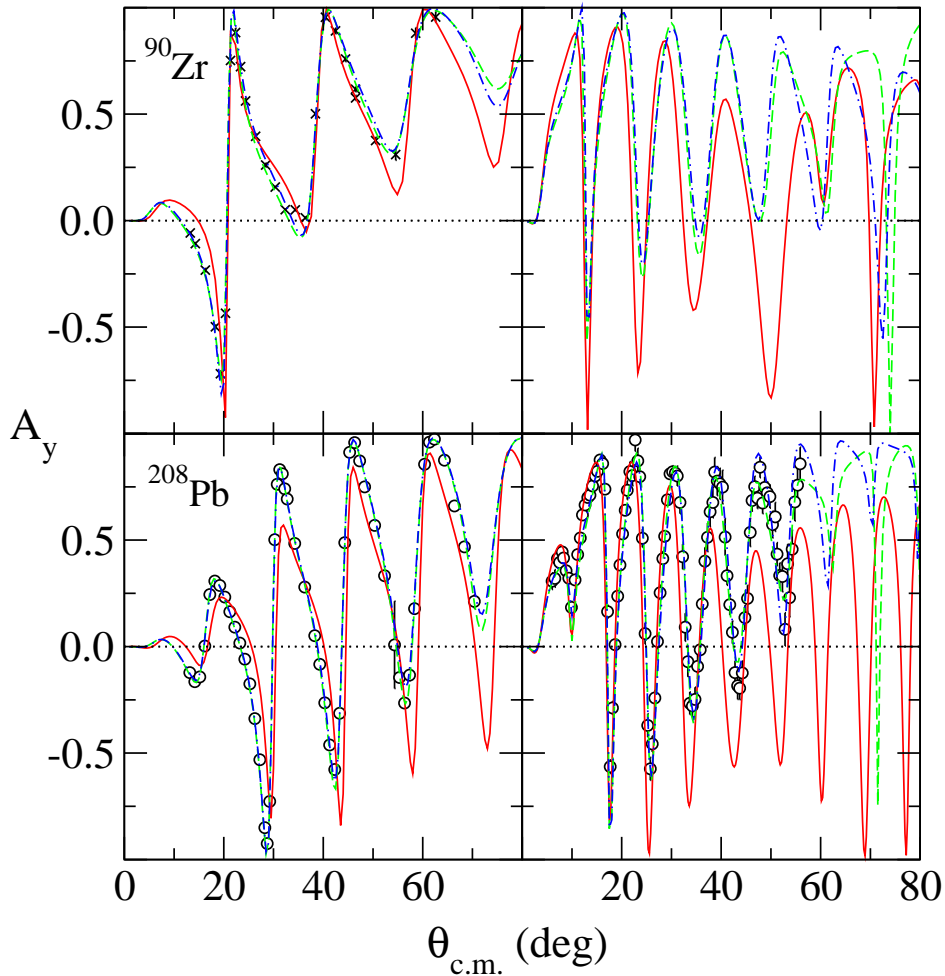}}
\caption{\label{ZPb-Ay}(Colour online)
Analyzing powers from 65 (left) and 200 (right) MeV proton elastic
scattering from ${}^{90}$Zr (top) and ${}^{208}$Pb (bottom).
Details are given in the text.}
\end{figure}

\begin{figure}
\scalebox{0.7}{\includegraphics{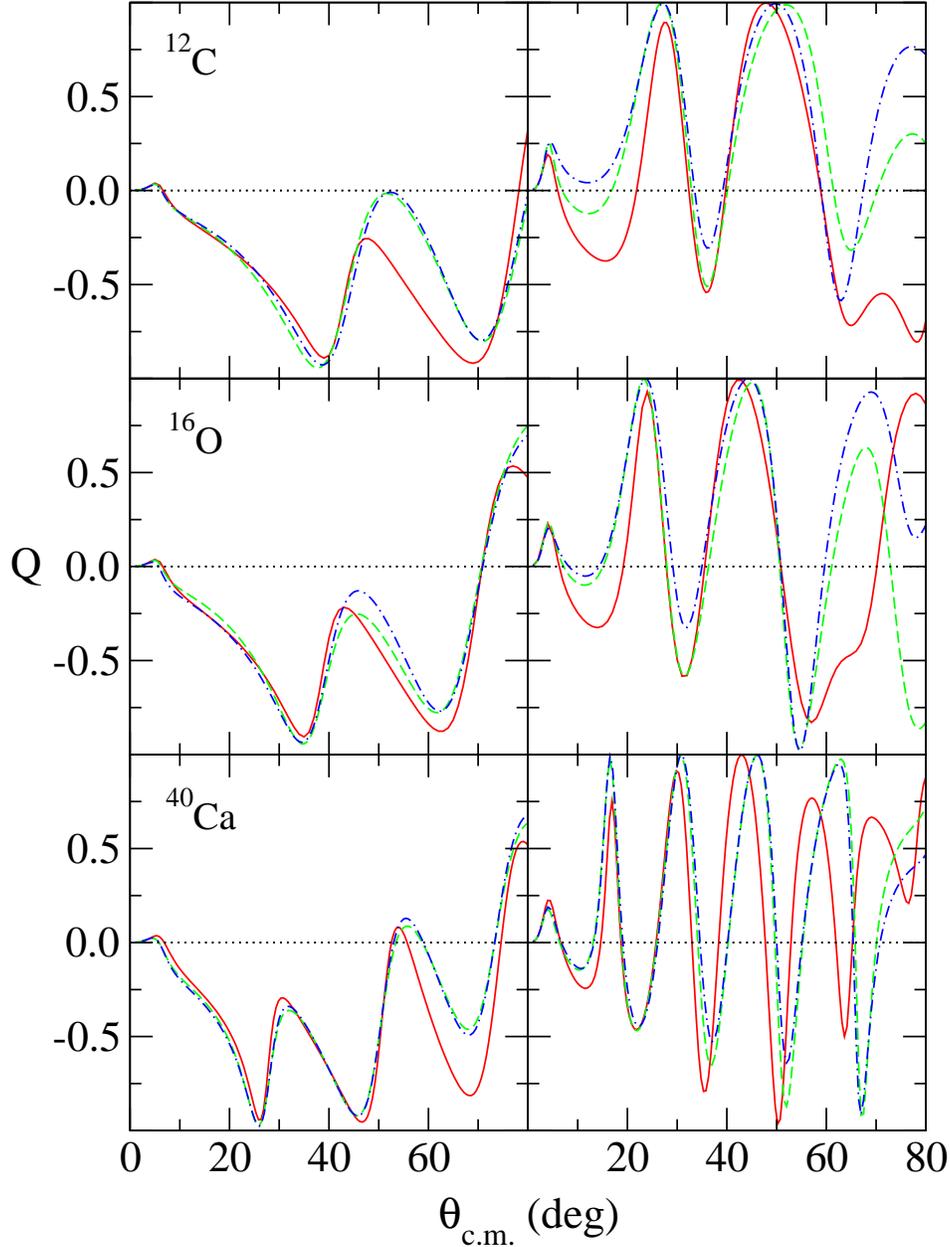}}
\caption{\label{CCa-Q}(Colour online)
Spin rotations from 65 (left) and 200 (right) MeV proton elastic
scattering from ${}^{12}$C (top), ${}^{16}$O (middle), and ${}^{40}$Ca (bottom).
Details are given in the text.}
\end{figure}

\begin{figure}
\scalebox{0.7}{\includegraphics{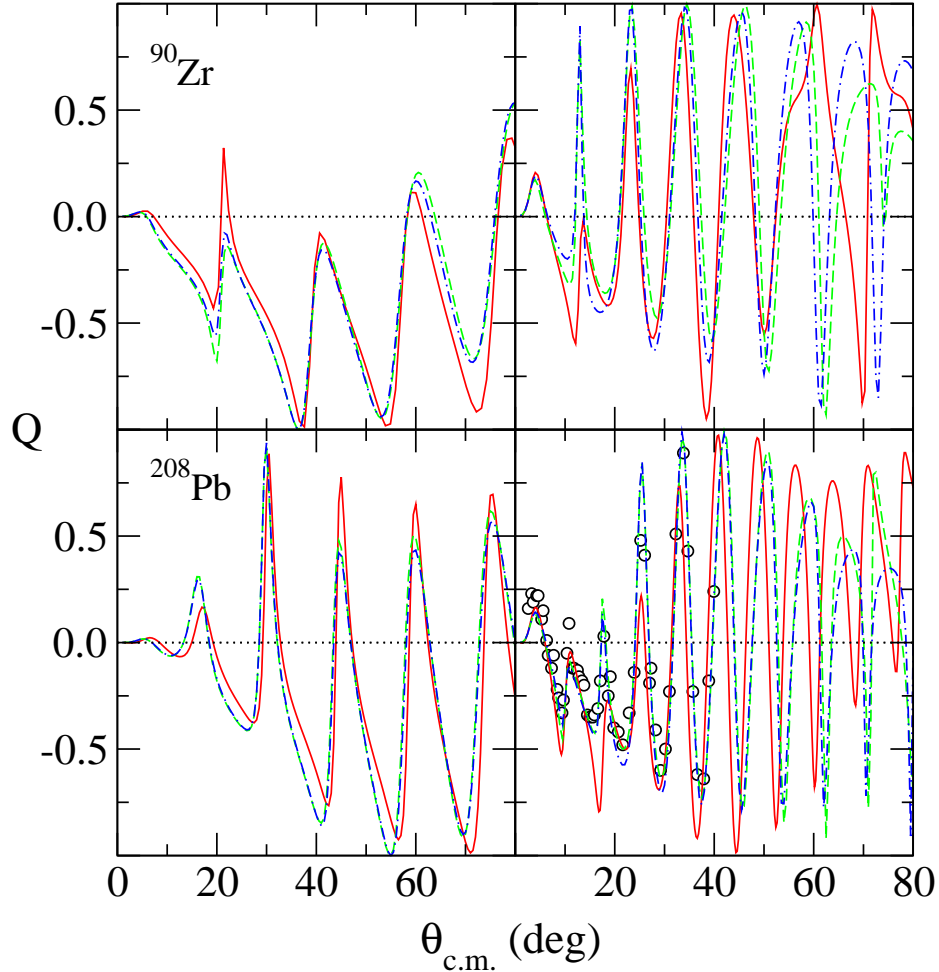}}
\caption{\label{ZPb-Q}(Colour online)
Spin rotations from 65 (left) and 200 (right) MeV proton elastic
scattering from ${}^{90}$Zr (top) and ${}^{208}$Pb (bottom).
Details are given in the text.}
\end{figure}

  The comparisons between data and calculated results for the spin observables
are displayed in Figs.~\ref{CCa-Ay}, \ref{ZPb-Ay}, \ref{CCa-Q}, and \ref{ZPb-Q}.
In the first two results for analyzing powers $A_y$ are displayed, while in the 
latter, two spin polarizations $Q$ are presented.    The results for 65~MeV and 
200~MeV  are  shown  on the left and right respectively in each figure.     The 
available data are presented by the diverse symbols in each diagram  while  the
calculated results are displayed by the curves.    The notation is as specified
previously.  In Fig.~\ref{CCa-Ay}, the analyzing powers from elastic scattering
of 65~MeV and 200~MeV protons from ${}^{12}$C, ${}^{16}$O, and ${}^{40}$Ca  are
shown.          Those  from  ${}^{90}$Zr  and  ${}^{208}$Pb  are  presented  in 
Fig.~\ref{ZPb-Ay}.  Once more the global DP calculations lead to very good fits 
to most of these data.

Disparities between the $EDAI$ and $EDAD3$ results,   and between those results
and data,    are  evident  at  large  scattering  angles for the ${}^{12}$C and 
${}^{16}$O cases.     But, as noted before, those regions involve high momentum
transfer and very small cross-section values;  features that are hardest to fit
within a global parameter scheme, let alone with a $g$-folding approach.     Of
course,   if each target is treated independently then it should be possible to 
find better phenomenological fits to these higher momentum transfer data.    As
with the cross sections, the predictions obtained with the $g$-folding approach,
while not reproducing the data as well as the global DP,    still are
quite good in comparison with the data.      In  some  cases,  the  $g$-folding
predictions are very good fits.   Overall then,  the  ingredients  used  in the 
$g$-folding method can be considered quite realistic.  

 The spin rotations $Q$  resulting from the three calculations are  compared in 
Figs.~\ref{CCa-Q} and \ref{ZPb-Q}.   There are very little data available,  but 
that for ${}^{208}$Pb is shown.   All  three calculated results agree very well 
with that (limited) data set.   For the other nuclei, the global DP results are
quite similar, diverging from each similarly   to  the  differences between the 
analyzing power results.  They also differ from the $g$-folding results in like
fashion.   As with the comparisons of analyzing powers, the spin rotations from 
all three calculations are in quite good agreement  structurally  to  over 
40$^\circ$ in the center of mass.

\subsection{Total reaction and total cross section results}

\subsubsection{Proton total reaction cross sections}

In this subsection predictions of the total reaction cross sections for proton 
scattering are displayed. Results for energies to 800~MeV for the five  nuclei 
considered are compared against data;  the latter taken from  many sources  as 
listed in Table~\ref{exptab}.  
\begin{table}
\begin{ruledtabular}
\caption{\label{exptab} 
Sources of proton reaction cross-section data.}
\begin{tabular}{cl}
 Nucleus   & Proton references (in year order)\\
\hline
$^{12}$C   &
\cite{Mi54}, \cite{Ca54}, \cite{Bu59}, \cite{Go59},
\cite{Me60}, \cite{Jo61}, \cite{Go62},
\cite{Gi64}, \cite{Ma64}, \cite{Ki66}, \cite{Me71},
\cite{Re72}, \cite{Mc74}, \cite{Sl75},
\cite{In99}\\
$^{16}$O   & \cite{Ch67}, \cite{Re72}, \cite{Ca75}\\
$^{40}$Ca  &
\cite{Jo61}, \cite{Tu64}, \cite{Ki66}, \cite{Di70}, \cite{Ca75}, 
\cite{In99}\\
$^{90}$Zr  & \cite{Wi63}, \cite{Ki66}, \cite{Di67}, \cite{Me71}\\
$^{208}$Pb & \cite{Ca54}, \cite{Go59}, \cite{Me60}, \cite{Go62},
\cite{Tu64}, \cite{Po65}, \cite{Ki66}, \cite{Me71}, \cite{Re72},
\cite{Mo73}, \cite{Ca75}, \cite{In99}\\
\end{tabular}
\end{ruledtabular}
\end{table}

\begin{figure}
\scalebox{0.8}{\includegraphics{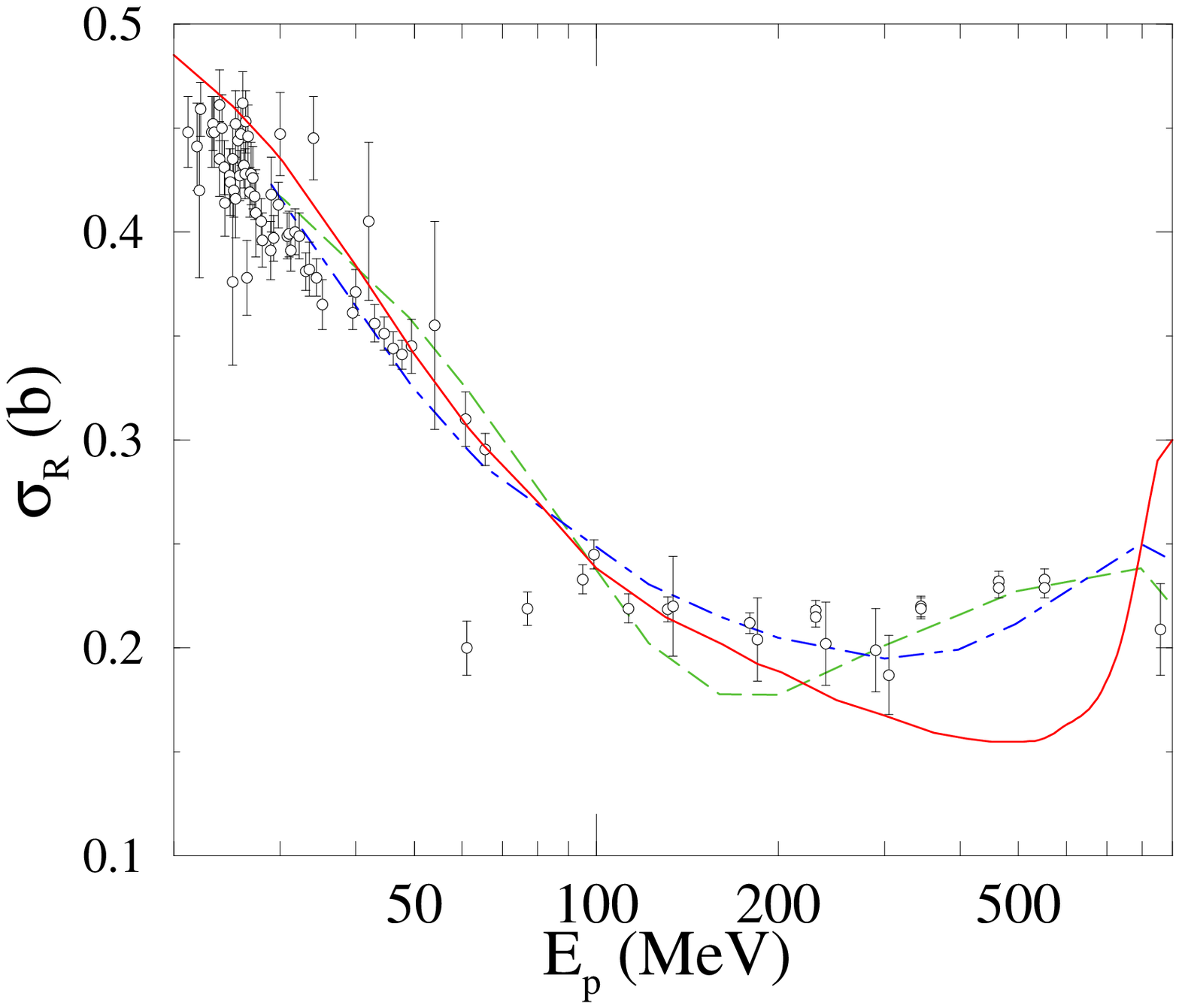}}
\caption{\label{comp-fig1} (Colour online)
Energy variation of $\sigma_R$ for
proton scattering from $^{12}$C. The curves are identified in the text while
data references are as given in Table~\ref{exptab}.}
\end{figure}

The results for scattering from $^{12}$C are presented in Fig.~\ref{comp-fig1}. 
Total proton reaction cross sections are  shown  for a projectile energy range 
from 20 MeV to 900 MeV.    Although  experimental  data  exist  to  much lower 
energies, any effective mean field prescription for  the  optical potential at 
such low energies is not appropriate. However, in a recent  paper~\cite{Am03}, 
a multi-channel  algebraic  scattering (MCAS) theory  has  been developed with 
which the low energy range should be covered.  This MCAS approach lends itself 
to specification of  the  dynamic  polarization  potential  resulting  from  a 
summation over all diverse channel functions allowed in that theory.   But the 
resultant optical potential then is extremely non-local and energy  dependent; 
as it need be not only to produce a smooth energy varying  elastic  scattering 
cross section but also to give the sharp and broad resonances readily observed 
at low ($< 20$ MeV) energies~\cite{Am03}.  The  low  level  densities  in most 
spectra to $\sim 20$ MeV is  not  a  condition  conducive to confidence in the 
$g$-folding or any  mean  field  optical  model potentials.   Nonetheless  for 
${}^{12}$C, there are many data points at energies between 20 and 40 MeV and a 
significant number thereafter to 500 MeV.  Above that there are  only a few. 

   In Fig.~\ref{comp-fig1},  the  data  to  300 MeV are well reproduced by the 
$g$-folding model results.    But  for  the  higher energies,  the $g$-folding 
approach is inadequate; seriously under-predicting the data.   That is the case 
for all five targets emphasizing the energy regime limit one need remember for
this microscopic Schr\"odinger approach to scattering.   On the other hand the 
reaction cross sections determined with  both the  $EDAI$  and  $EDAD3$ global 
potentials reflect the data very well falling within the uncertainties at most 
energies.  Some data, notably at 61~MeV \cite{Me60} and at  77~MeV \cite{Go62} 
are in disagreement with all three  calculated  results.   However,  they  are 
exceptional points in that they also disagree with  empirical  values taken at 
nearby energies.    Menet~{\it et al.}~\cite{Me71}  argue  that  a much larger 
systematic  error  should  have  been  used  in the data processing of earlier 
experiments that lead to the exceptional points.   Such  also  occur with data 
taken on other targets.

\begin{figure}
\scalebox{0.8}{\includegraphics{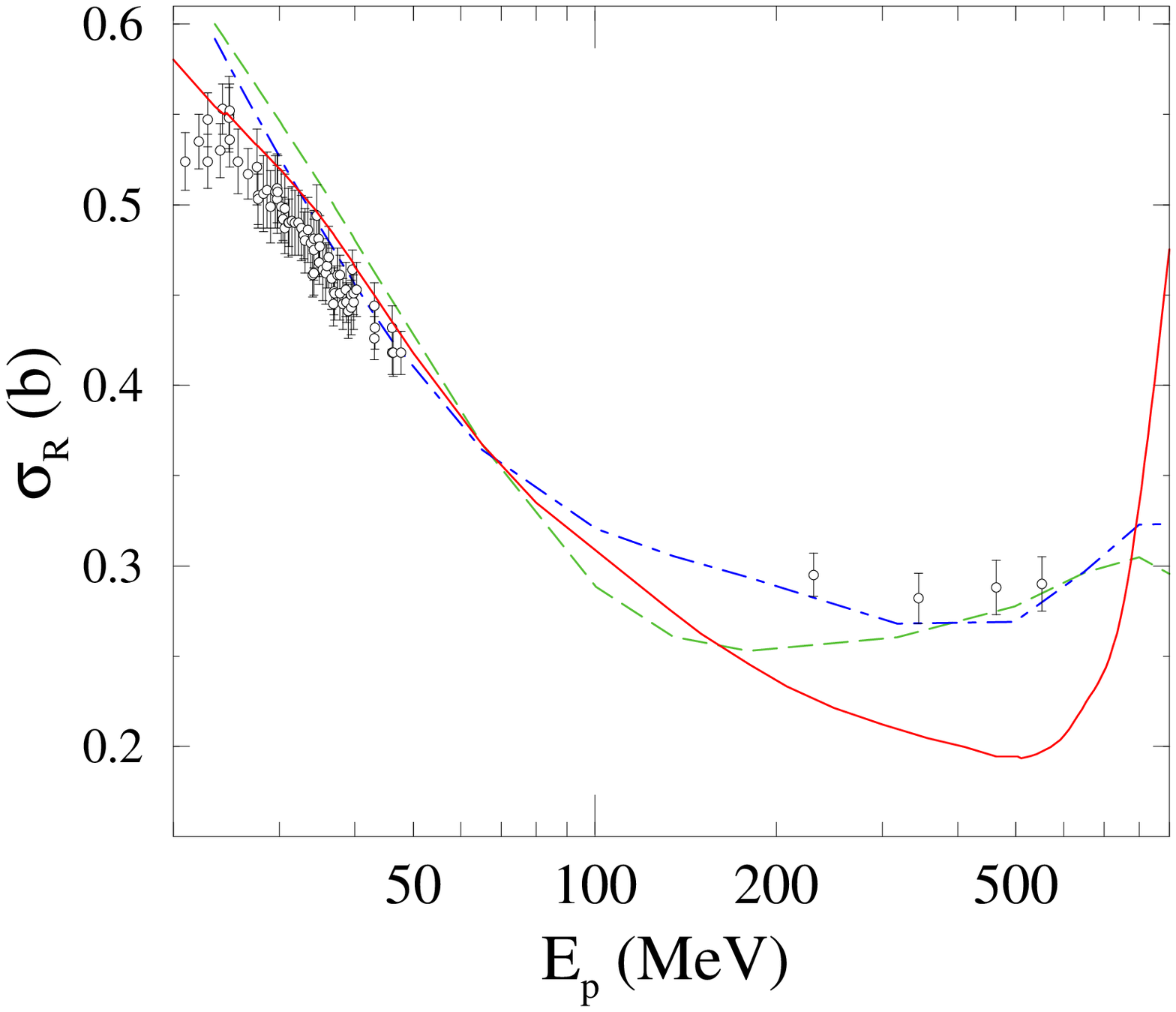}}
\caption{\label{comp-fig2}(Colour online)
 Energy variation of $\sigma_R$ for proton 
scattering from  $^{16}$O.  The notation is as in the text.}
\end{figure}

Predictions for proton scattering from $^{16}$O are compared with data
in Fig.~\ref{comp-fig2}.
There is no p-$^{16}$O total reaction cross section data for energies  between 
50~MeV and 200~MeV. Nonetheless the $g$-folding potential predictions to 50~MeV 
are in very good agreement with data for those energies.   But as with the case 
of ${}^{12}$C, the $g$-folding approach under-predicted the four data  points in
the energy range 200~MeV and 600~MeV.       These  high  energy  data  are well
replicated by the global DP potentials however,   and  by  the $EDAD3$ model in 
particular. That model also gave results that compare well the data  at  lower
energies while the $EDAI$ result over-predicted them.

\begin{figure}
\scalebox{0.8}{\includegraphics{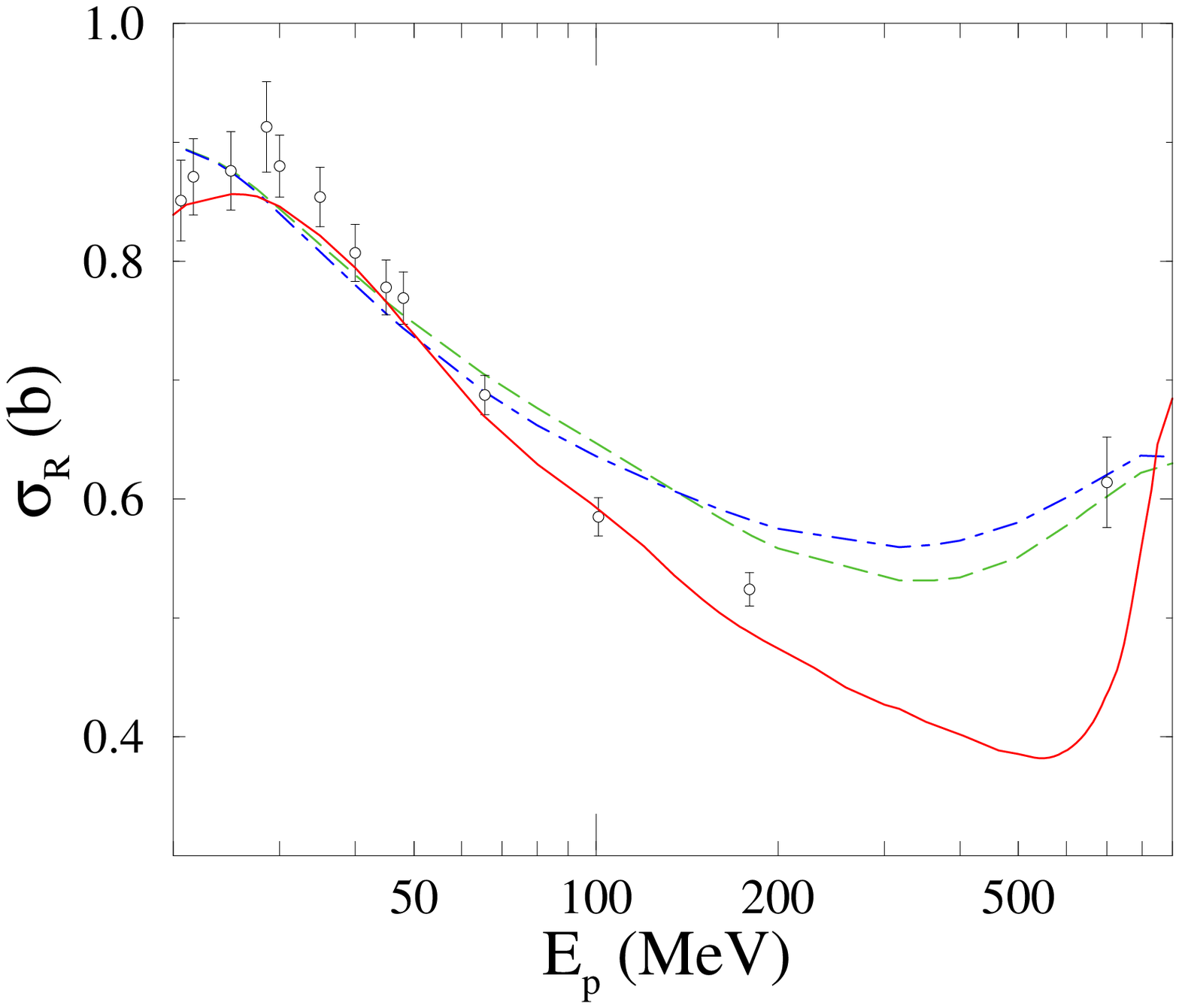}}
\caption{\label{comp-fig3} (Colour online)
Energy dependencies of $\sigma_R$ for 
proton scattering from $^{40}$Ca. The notation is as in the text.}
\end{figure}

 Predictions  for  proton  scattering from $^{40}$Ca are compared with  data in 
Fig.~\ref{comp-fig3}.  There are relatively few data points in the energy range 
with but one (at 700 MeV) above 180 MeV.   Again the predictions made using the 
$g$-folding potentials are in very good agreement with the data for energies to 
180 MeV. But the microscopic approach underestimates the datum at 700 MeV. Both 
the global  DP  ($EDAI$ and $EDAD3$)  models  replicate  the  data  well at the 
energies to   65~MeV,  overestimate the data at the energies  near  100~MeV and 
180~MeV, and reproduce very well the datum at 700~MeV.   

\begin{figure}
\scalebox{0.8}{\includegraphics{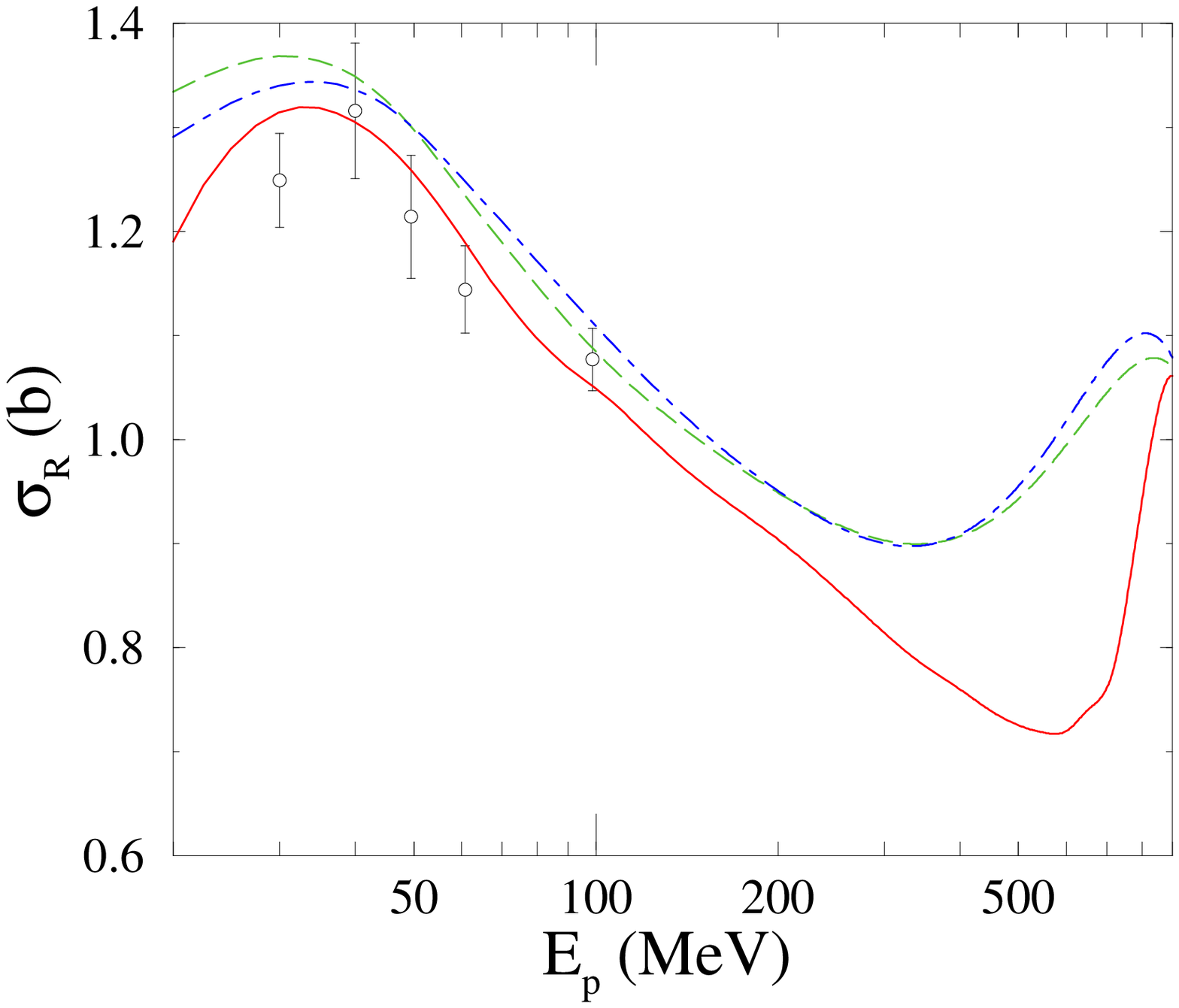}}
\caption[]{\label{comp-fig4}(Colour online) 
Energy dependencies of $\sigma_R$
for proton scattering from $^{90}$Zr.  Basic notation is as for
Fig.~\ref{comp-fig1}.}
\end{figure}

  In Fig.~\ref{comp-fig4}, we present the data and our predictions of the total 
reaction cross sections for proton scattering from $^{90}$Zr.   There  is  very 
little data with which to compare, nevertheless the $g$-folding potential gives 
results in very good agreement with the five actual data points  in  the energy
range 30 to 100~MeV.   On  the  other  hand  both  global DP model calculations
slightly overestimate that data set.  Again however, it is in the higher energy
ranges that the $g$-folding and Dirac model predictions  diverge,  as  we  have
come to expect.

\begin{figure}
\scalebox{0.8}{\includegraphics{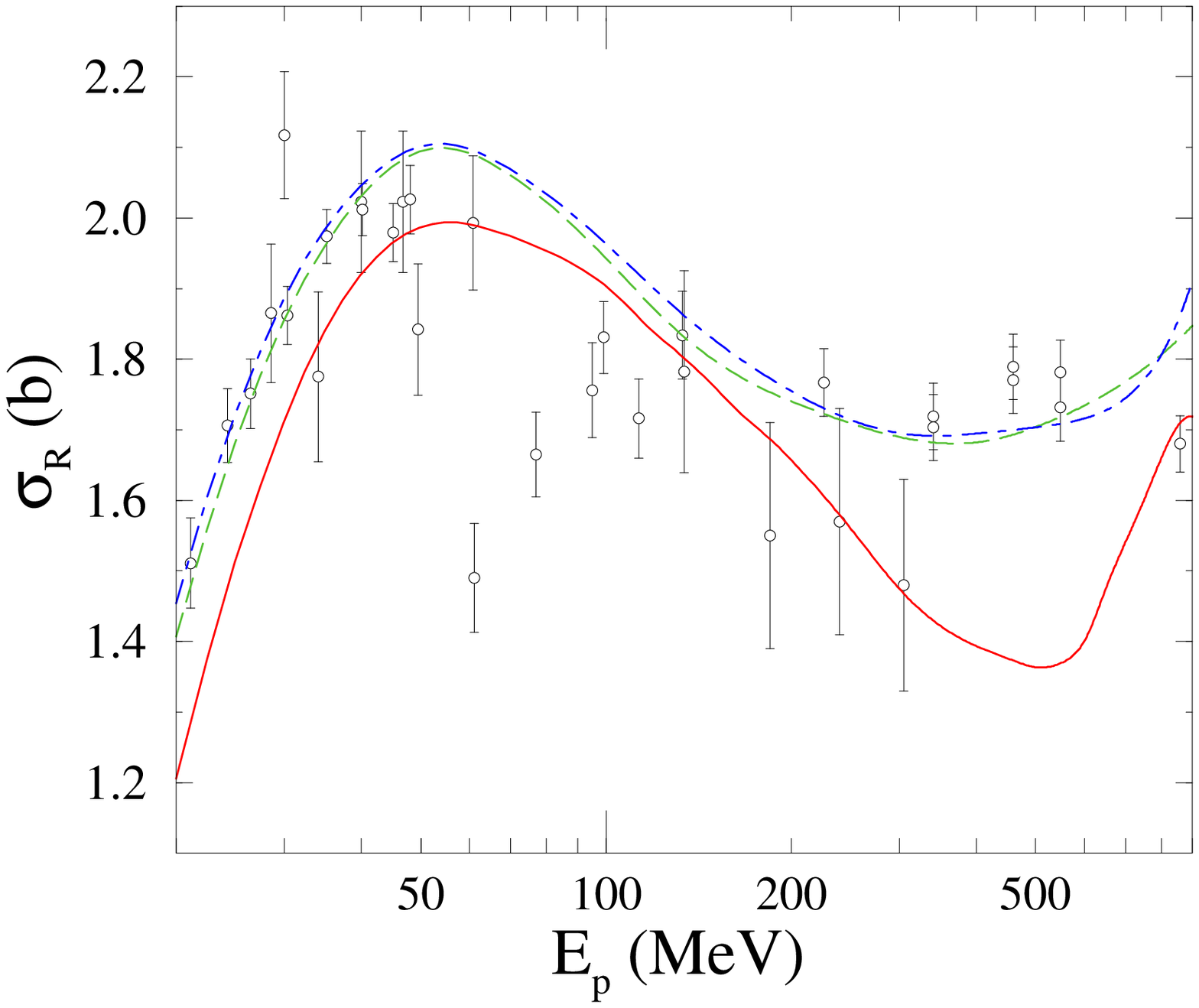}}
\caption[]{\label{comp-fig5}(Colour online) 
Energy dependencies of $\sigma_R$
for proton scattering from $^{208}$Pb.  The notation is as in the text.}
\end{figure}
 In Fig.~\ref{comp-fig5}, we compare the calculated total proton reaction cross 
sections  from   $^{208}$Pb  with  data  taken  from  the  references listed in 
Table~\ref{exptab}.  For  this  nucleus, the data are quite scattered with some 
being quite  disparate.    In  this  case  the  $g$-folding  model calculations
underestimate data at energies between 20~MeV and 40~MeV  but  agree  well with
the data taken at energies thereafter to 300~MeV.  On  the  other  hand,   both
$EDAI$ and $EDAD3$ global DP models reproduce a select set of the data over the
whole energy range quite well.  As with ${}^{12}$C, exceptional data points are
found near 30~MeV, 61~MeV and 77~MeV.  In this case, other data from $^{208}$Pb
taken at 60.8~MeV~\cite{Me71} and  65.5~MeV~\cite{In99} give different  results
and in fact are values of reaction cross sections that are in accord  with  our
model predictions.

\subsubsection{Energy variations of neutron total cross sections}

  Accurate  measurements  of neutron total reaction cross sections are far more 
difficult to achieve than their proton counterparts. Indeed usually those cross 
sections  are  obtained  by  subtracting  the elastic from the total scattering 
cross section.   While  both  of those cross sections can be measured with some
accuracy, the  subtraction  of  two large  numbers with attendant uncertainties
compounds uncertainty in the result.   We  note,  however,  that a new Japanese
technique,  for  measuring  neutron  total  reaction  cross  sections utilizing 
in-beam and out-beam methods similar to those used in proton scattering,  shows
promise.   Nevertheless, herein, we concentrate on analyses of total scattering
cross-section data.

 The  data  we  have  chosen to analyze have been taken from a recent survey by 
Abfalterer~{\em et al.}~\cite{Ab01}.  That  survey  includes  data  measured at 
LANSCE  that  are  supplementary  and  additional to those published earlier by 
Finlay~{\em et al.}~\cite{Fi93}.      For  comparison  with  that  recent  data 
compilation, we calculate neutron  total  scattering  cross  sections  for  the
five   nuclei  of  interest;  $^{12}$C,  $^{16}$O,  $^{40}$Ca,  $^{90}$Zr,  and
$^{208}$Pb .

\begin{figure}
\scalebox{0.8}{\includegraphics{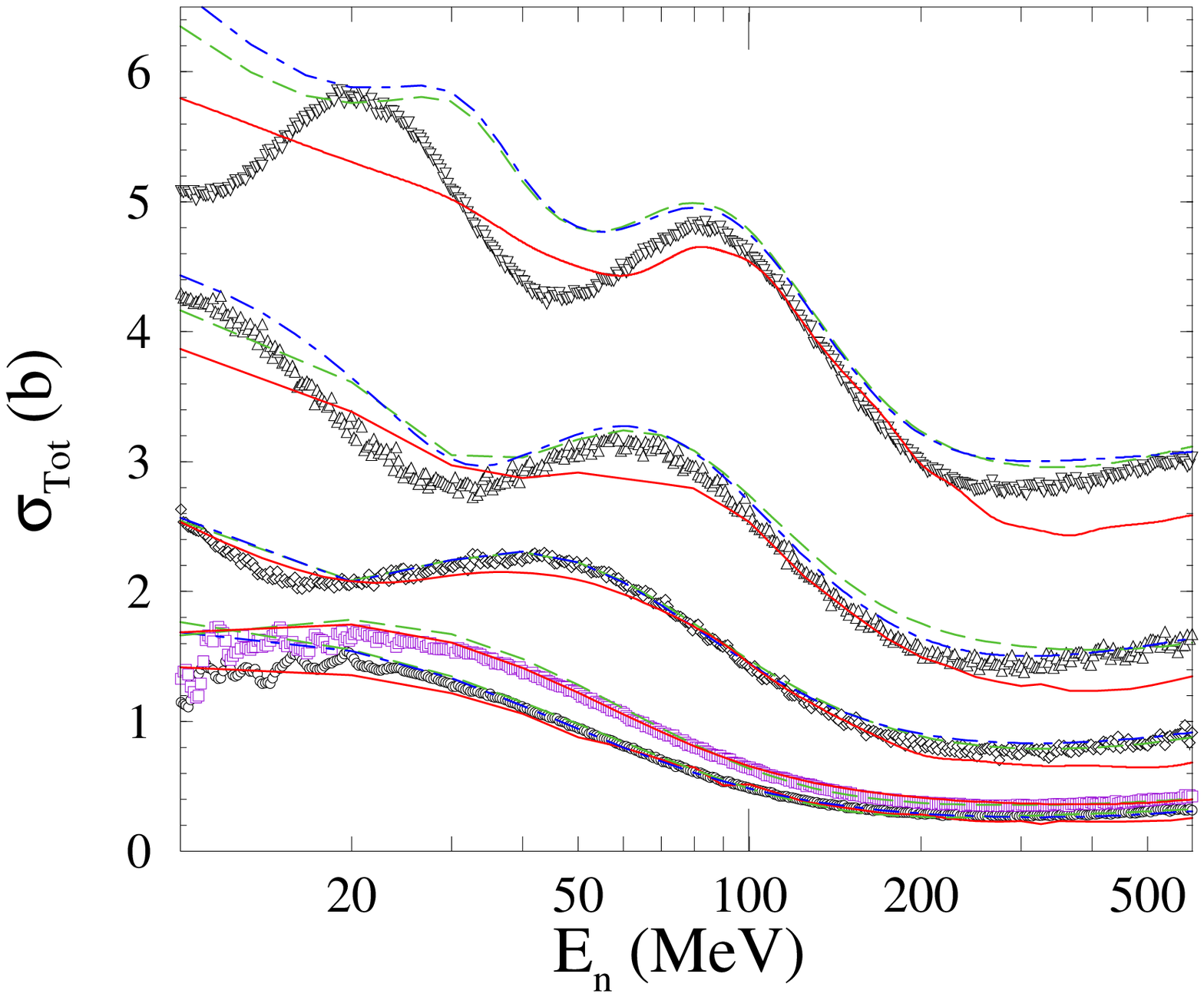}}
\caption{\label{comp-fig6}(Colour online)
Total cross sections for neutrons scattered from 
$^{12}$C, $^{16}$O, $^{40}$Ca, $^{90}$Zr, and $^{208}$Pb. The data are
those of Abfalterer {\em et al.} \cite{Ab01}; for $^{12}$C the data
correspond to scattering from natural carbon.}
\end{figure}

In forming neutron optical potentials by the $g$-folding method,  we  have used
the same structure models and effective $NN$ interactions chosen  to create the
proton optical potentials whose results were  presented  and  discussed  above.
Likewise global DP model potentials for neutrons were formed using  the  $EDAI$ 
and $EDAD3$ parameter sets that have been used to get proton scattering results, 
and with the Coulomb potential set to zero,  generated  results  displayed  and
discussed.  No Coulomb energy shift has been made in those calculations.   Thus 
all of the results shown in Fig.~\ref{comp-fig6} are predictions. 
For $^{12}$C, neutron total cross sections are well predicted by all three model
calculations. The $EDAI$ and $EDAD3$ potentials of the global DP give very good 
results for energies above 20~MeV, while the $g$-folding model calculations  do 
so  for  energies  between  50~MeV and 300~MeV.   Outside  of  that  range  the 
$g$-folding results under-predict data. Below 20~MeV in this case, as well as for
${}^{16}$O which is considered next, resonance effects are very evident.   Such 
are not encompassed in the mean field models  we consider.   For  $^{16}$O, the 
model results follow a similar pattern as for ${}^{12}$C.     Global DP results 
give good fits to the data  as does the $g$-folding model but only for energies 
above $\sim 70$~MeV.  The $EDAD3$ model predicts well data for the whole energy 
range while the $EDAI$ model  slightly  over-predicts  data  between 20~MeV and 
60~MeV. 

For the three heavier nuclei, ${}^{40}$Ca, ${}^{90}$Zr, and  ${}^{208}$Pb,  the 
total cross sections down to 10~MeV do not show significant resonance  features 
but there is the onset of a Ramsauer-like effect which accentuates with  target 
mass.    None  of  the models predict such large scale oscillations adequately, 
though  with  improved  structure  details,  as  used  in  the calculations for 
scattering  from  ${}^{208}$Pb,  those  oscillations  are  more noticeable when 
compared   to   results   from   use  of   a  simple   oscillator   model   of 
structure~\cite{Am02}.    The  results  found  for  these  three  nuclei show a 
degradation in the quality of the predictions from all three model calculations. 
For ${}^{40}$Ca,  the  comparisons are still similar to what is noticed for the 
lighter mass targets, though the $g$-folding model predictions above 300 MeV are 
poorer.    For  $^{90}$Zr and $^{208}$Pb all three model results clearly depart 
from what is observed.    However,  the $g$-folding results trend quite well as  
average results upon which a Ramsauer effect may be superimposed~\cite{De04}.

\section{Conclusions}

  A  microscopic  coordinate  space  Schr\"odinger  model  of  nucleon-nucleus 
interactions and global DP model for the same systems, have been used to predict 
proton   and   neutron  scattering  cross  sections; both angular and integral.  
Cross-section  data  and spin observables have been well described for energies 
at which the diverse models are valid.   That  includes  a substantial range of 
energies in common for both approaches.  

  Our first study was of  angular  dependent  observables;  differential cross 
sections and spin observables (analyzing powers and spin rotations) for elastic 
proton scattering at 65~MeV and at 200~MeV.  Both methods of analysis gave good 
reproduction of measured data.               The two models, with the global DP
potentials defined  from  the  parameter specifications  in the recent study of 
their energy and mass dependences~\cite{Co93}, were used to predict proton total
reaction  and  neutron  total  cross  sections  in scattering from five nuclei, 
${}^{12}$C, ${}^{16}$O, ${}^{40}$Ca, ${}^{90}$Zr, and ${}^{208}$Pb. Both models 
gave good predictions of that data (in the energy regimes for which the models 
are appropriate)  with  the  largest  discrepancies  reflecting  the 
Ramsauer effect in neutron total cross sections.  

   Fundamentally,  our  Dirac and  Schr\"odinger approaches are different.  The 
microscopic   Schr\"odinger   model   incorporates  all  the  dominant  medium 
modifications in the optical potential, without significant approximation,  by 
using a realistic ground state density to give a reasonable specification of all
terms in the optical potential.   The  global  DP  approach  provides a natural 
specification of such  terms  in  local  equivalent  form.    Nonetheless,  the 
comparisons obtained in this work are exemplary of the dilemma in  judging  the 
relative merits of relativistic {\em vs.} non-relativistic approaches  for  the 
analyses of intermediate energy nucleon-nucleus scattering. That the two models 
give results that agree so well for the differential cross sections as well  as 
for both spin observables  and  for  both  projectiles gives confidence in  the
reality of  calculated results at the two energies we considered.   As 
has  been  speculated~\cite{Ra92}  however,  the  answer  throughout the energy 
regime  may  lie  in  QCD-based  models  of  nuclear scattering systems.   
Indeed the work of Cohen {\em et al.}~\cite{Co91} shows evidence that the large
scalar and vector fields of Dirac Phenomenology may be related to quark degrees
of freedom in the nucleon.
However, such QCD-based
theoretical  models  and  concomitant  experiments  will  require  simultaneous 
treatment of proton and neutron scattering in order to be complete.  Sadly, the 
latter's experimental database currently is lacking.


\begin{acknowledgments}
This research was supported  by the National Science Foundation under 
Grant No. 0098645 and by a research grant from the Australian Research Council.
\end{acknowledgments}

\bibliography{Gfold+Dirac}

\begin{thebibliography}{57}
\expandafter\ifx\csname natexlab\endcsname\relax\def\natexlab#1{#1}\fi
\expandafter\ifx\csname bibnamefont\endcsname\relax
  \def\bibnamefont#1{#1}\fi
\expandafter\ifx\csname bibfnamefont\endcsname\relax
  \def\bibfnamefont#1{#1}\fi
\expandafter\ifx\csname citenamefont\endcsname\relax
  \def\citenamefont#1{#1}\fi
\expandafter\ifx\csname url\endcsname\relax
  \def\url#1{\texttt{#1}}\fi
\expandafter\ifx\csname urlprefix\endcsname\relax\def\urlprefix{URL }\fi
\providecommand{\bibinfo}[2]{#2}
\providecommand{\eprint}[2][]{\url{#2}}

\bibitem[{\citenamefont{Ray et~al.}(1992)\citenamefont{Ray, Hoffmann, and
  Coker}}]{Ra92}
\bibinfo{author}{\bibfnamefont{L.}~\bibnamefont{Ray}},
  \bibinfo{author}{\bibfnamefont{G.~W.} \bibnamefont{Hoffmann}},
  \bibnamefont{and} \bibinfo{author}{\bibfnamefont{W.~R.} \bibnamefont{Coker}},
  \bibinfo{journal}{Phys. Rep.} \textbf{\bibinfo{volume}{212}},
  \bibinfo{pages}{223} (\bibinfo{year}{1992}), \bibinfo{note}{and references
  cited therein}.

\bibitem[{\citenamefont{Amos et~al.}(2000)\citenamefont{Amos, Dortmans, von
  Geramb, Karataglidis, and Raynal}}]{Am00}
\bibinfo{author}{\bibfnamefont{K.}~\bibnamefont{Amos}},
  \bibinfo{author}{\bibfnamefont{P.~J.} \bibnamefont{Dortmans}},
  \bibinfo{author}{\bibfnamefont{H.~V.} \bibnamefont{von Geramb}},
  \bibinfo{author}{\bibfnamefont{S.}~\bibnamefont{Karataglidis}},
  \bibnamefont{and} \bibinfo{author}{\bibfnamefont{J.}~\bibnamefont{Raynal}},
  \bibinfo{journal}{Adv. in Nucl. Phys.} \textbf{\bibinfo{volume}{25}},
  \bibinfo{pages}{275} (\bibinfo{year}{2000}), \bibinfo{note}{and references
  cited therein}.

\bibitem[{\citenamefont{Serot and Walecka}(1986)}]{Se86}
\bibinfo{author}{\bibfnamefont{B.~D.} \bibnamefont{Serot}} \bibnamefont{and}
  \bibinfo{author}{\bibfnamefont{J.~D.} \bibnamefont{Walecka}},
  \bibinfo{journal}{Adv. Nucl. Phys.} \textbf{\bibinfo{volume}{16}},
  \bibinfo{pages}{1} (\bibinfo{year}{1986}), \bibinfo{note}{and references
  cited therein}.

\bibitem[{\citenamefont{Cooper et~al.}(1993)\citenamefont{Cooper, Hama, Clark,
  and Mercer}}]{Co93}
\bibinfo{author}{\bibfnamefont{E.~D.} \bibnamefont{Cooper}},
  \bibinfo{author}{\bibfnamefont{S.}~\bibnamefont{Hama}},
  \bibinfo{author}{\bibfnamefont{B.~C.} \bibnamefont{Clark}}, \bibnamefont{and}
  \bibinfo{author}{\bibfnamefont{R.~L.} \bibnamefont{Mercer}},
  \bibinfo{journal}{Phys. Rev. C} \textbf{\bibinfo{volume}{47}},
  \bibinfo{pages}{297} (\bibinfo{year}{1993}), \bibinfo{note}{and references
  cited therein.}

\bibitem[{\citenamefont{Kozack and Madland}(1989)}]{Ko89}
\bibinfo{author}{\bibfnamefont{R.}~\bibnamefont{Kozack}} \bibnamefont{and}
  \bibinfo{author}{\bibfnamefont{D.~G.} \bibnamefont{Madland}},
  \bibinfo{journal}{Phys. Rev. C} \textbf{\bibinfo{volume}{39}},
  \bibinfo{pages}{1461} (\bibinfo{year}{1989}).

\bibitem[{\citenamefont{Deb et~al.}(2001)\citenamefont{Deb, Amos, Karataglidis,
  Chadwick, and Madland}}]{De01}
\bibinfo{author}{\bibfnamefont{P.~K.} \bibnamefont{Deb}},
  \bibinfo{author}{\bibfnamefont{K.}~\bibnamefont{Amos}},
  \bibinfo{author}{\bibfnamefont{S.}~\bibnamefont{Karataglidis}},
  \bibinfo{author}{\bibfnamefont{M.~B.} \bibnamefont{Chadwick}},
  \bibnamefont{and} \bibinfo{author}{\bibfnamefont{D.~G.}
  \bibnamefont{Madland}}, \bibinfo{journal}{Phys. Rev. Lett.}
  \textbf{\bibinfo{volume}{86}}, \bibinfo{pages}{3248} (\bibinfo{year}{2001}).

\bibitem[{\citenamefont{Brown}(2000)}]{Br00}
\bibinfo{author}{\bibfnamefont{B.~A.} \bibnamefont{Brown}},
  \bibinfo{journal}{Phys. Rev. Lett.} \textbf{\bibinfo{volume}{85}},
  \bibinfo{pages}{5296} (\bibinfo{year}{2000}).

\bibitem[{\citenamefont{Friedman and Pandharipande}(1981)}]{Fr81}
\bibinfo{author}{\bibfnamefont{B.}~\bibnamefont{Friedman}} \bibnamefont{and}
  \bibinfo{author}{\bibfnamefont{V.~R.} \bibnamefont{Pandharipande}},
  \bibinfo{journal}{Nucl. Phys.} \textbf{\bibinfo{volume}{A361}},
  \bibinfo{pages}{502} (\bibinfo{year}{1981}).

\bibitem[{\citenamefont{Clark et~al.}(2003)\citenamefont{Clark, Kerr, and
  Hama}}]{Cl03}
\bibinfo{author}{\bibfnamefont{B.~C.} \bibnamefont{Clark}},
  \bibinfo{author}{\bibfnamefont{L.~J.} \bibnamefont{Kerr}}, \bibnamefont{and}
  \bibinfo{author}{\bibfnamefont{S.}~\bibnamefont{Hama}},
  \bibinfo{journal}{Phys. Rev. C} \textbf{\bibinfo{volume}{67}},
  \bibinfo{pages}{054605} (\bibinfo{year}{2003}), \bibinfo{note}{and references
  cited therein}.

\bibitem[{\citenamefont{Karataglidis and Madland}(2001)}]{Ka01}
\bibinfo{author}{\bibfnamefont{S.}~\bibnamefont{Karataglidis}}
  \bibnamefont{and} \bibinfo{author}{\bibfnamefont{D.~G.}
  \bibnamefont{Madland}} (\bibinfo{year}{2001}),
  \bibinfo{note}{nucl-th/0103048}.

\bibitem[{\citenamefont{Koning and Delaroche}(2003)}]{Ko03}
\bibinfo{author}{\bibfnamefont{A.~J.} \bibnamefont{Koning}} \bibnamefont{and}
  \bibinfo{author}{\bibfnamefont{J.~P.} \bibnamefont{Delaroche}},
  \bibinfo{journal}{Nucl. Phys.} \textbf{\bibinfo{volume}{A713}},
  \bibinfo{pages}{231} (\bibinfo{year}{2003}).

\bibitem[{\citenamefont{Machleidt et~al.}(1987)\citenamefont{Machleidt,
  Holinde, and Elster}}]{Ma87}
\bibinfo{author}{\bibfnamefont{R.}~\bibnamefont{Machleidt}},
  \bibinfo{author}{\bibfnamefont{K.}~\bibnamefont{Holinde}}, \bibnamefont{and}
  \bibinfo{author}{\bibfnamefont{C.}~\bibnamefont{Elster}},
  \bibinfo{journal}{Phys. Rep.} \textbf{\bibinfo{volume}{149}},
  \bibinfo{pages}{1} (\bibinfo{year}{1987}).

\bibitem[{\citenamefont{Raynal}(1999)}]{Ra99}
\bibinfo{author}{\bibfnamefont{J.}~\bibnamefont{Raynal}},
  \emph{\bibinfo{title}{computer code dwba98}} (\bibinfo{year}{1999}),
  \bibinfo{note}{(NEA 1209/05)}.

\bibitem[{\citenamefont{Clark et~al.}(1985)\citenamefont{Clark, Hama,
  K{\"a}lbermann, Cooper, and Mercer}}]{Cl85}
\bibinfo{author}{\bibfnamefont{B.~C.} \bibnamefont{Clark}},
  \bibinfo{author}{\bibfnamefont{S.}~\bibnamefont{Hama}},
  \bibinfo{author}{\bibfnamefont{S.~G.} \bibnamefont{K{\"a}lbermann}},
  \bibinfo{author}{\bibfnamefont{E.~D.} \bibnamefont{Cooper}},
  \bibnamefont{and} \bibinfo{author}{\bibfnamefont{R.~L.}
  \bibnamefont{Mercer}}, \bibinfo{journal}{Phys. Rev. C}
  \textbf{\bibinfo{volume}{31}}, \bibinfo{pages}{694} (\bibinfo{year}{1985}).

\bibitem[{\citenamefont{Cooper and Jennings}(1988)}]{Co88}
\bibinfo{author}{\bibfnamefont{E.~D.} \bibnamefont{Cooper}} \bibnamefont{and}
  \bibinfo{author}{\bibfnamefont{B.~K.} \bibnamefont{Jennings}},
  \bibinfo{journal}{Nucl. Phys.} \textbf{\bibinfo{volume}{A483}},
  \bibinfo{pages}{601} (\bibinfo{year}{1988}).

\bibitem[{\citenamefont{Landau}(1990)}]{La90}
\bibinfo{author}{\bibfnamefont{R.~H.} \bibnamefont{Landau}},
  \emph{\bibinfo{title}{Quantum Mechanics II, {\it A Second Course in Quantum
  Theory}}} (\bibinfo{publisher}{Wiley}, \bibinfo{address}{New York},
  \bibinfo{year}{1990}).

\bibitem[{\citenamefont{Karataglidis et~al.}(1995)\citenamefont{Karataglidis,
  Dortmans, Amos, and de~Swiniarski}}]{Ka95}
\bibinfo{author}{\bibfnamefont{S.}~\bibnamefont{Karataglidis}},
  \bibinfo{author}{\bibfnamefont{P.~J.} \bibnamefont{Dortmans}},
  \bibinfo{author}{\bibfnamefont{K.}~\bibnamefont{Amos}}, \bibnamefont{and}
  \bibinfo{author}{\bibfnamefont{R.}~\bibnamefont{de~Swiniarski}},
  \bibinfo{journal}{Phys. Rev. C} \textbf{\bibinfo{volume}{52}},
  \bibinfo{pages}{861} (\bibinfo{year}{1995}).

\bibitem[{\citenamefont{Warburton and Brown}(1992)}]{Wa92}
\bibinfo{author}{\bibfnamefont{E.~K.} \bibnamefont{Warburton}}
  \bibnamefont{and} \bibinfo{author}{\bibfnamefont{B.~A.} \bibnamefont{Brown}},
  \bibinfo{journal}{Phys. Rev. C} \textbf{\bibinfo{volume}{46}},
  \bibinfo{pages}{923} (\bibinfo{year}{1992}).

\bibitem[{\citenamefont{Haxton and Johnson}(1990)}]{Ha90}
\bibinfo{author}{\bibfnamefont{W.~C.} \bibnamefont{Haxton}} \bibnamefont{and}
  \bibinfo{author}{\bibfnamefont{C.}~\bibnamefont{Johnson}},
  \bibinfo{journal}{Phys. Rev. Lett} \textbf{\bibinfo{volume}{65}},
  \bibinfo{pages}{1325} (\bibinfo{year}{1990}).

\bibitem[{\citenamefont{Cohen and Kurath}(1965)}]{Co65}
\bibinfo{author}{\bibfnamefont{S.}~\bibnamefont{Cohen}} \bibnamefont{and}
  \bibinfo{author}{\bibfnamefont{D.}~\bibnamefont{Kurath}},
  \bibinfo{journal}{Nucl. Phys.} \textbf{\bibinfo{volume}{73}},
  \bibinfo{pages}{1} (\bibinfo{year}{1965}).

\bibitem[{\citenamefont{Karataglidis et~al.}(2002)\citenamefont{Karataglidis,
  Amos, Brown, and Deb}}]{Ka02}
\bibinfo{author}{\bibfnamefont{S.}~\bibnamefont{Karataglidis}},
  \bibinfo{author}{\bibfnamefont{K.}~\bibnamefont{Amos}},
  \bibinfo{author}{\bibfnamefont{B.~A.} \bibnamefont{Brown}}, \bibnamefont{and}
  \bibinfo{author}{\bibfnamefont{P.~K.} \bibnamefont{Deb}},
  \bibinfo{journal}{Phys. Rev. C} \textbf{\bibinfo{volume}{65}},
  \bibinfo{pages}{044306} (\bibinfo{year}{2002}), \bibinfo{note}{and references
  cited therein}.

\bibitem[{\citenamefont{Ji and Wildenthal}(1989)}]{Ji89}
\bibinfo{author}{\bibfnamefont{X.}~\bibnamefont{Ji}} \bibnamefont{and}
  \bibinfo{author}{\bibfnamefont{B.~H.} \bibnamefont{Wildenthal}},
  \bibinfo{journal}{Phys. Rev. C} \textbf{\bibinfo{volume}{40}},
  \bibinfo{pages}{389} (\bibinfo{year}{1989}).

\bibitem[{\citenamefont{Noro et~al.}(1981)}]{No81}
\bibinfo{author}{\bibfnamefont{T.}~\bibnamefont{Noro}} \bibnamefont{et~al.},
  \bibinfo{journal}{Nucl. Phys.} \textbf{\bibinfo{volume}{A366}},
  \bibinfo{pages}{189} (\bibinfo{year}{1981}).

\bibitem[{\citenamefont{Sakaguchi et~al.}(1982)}]{Sa82}
\bibinfo{author}{\bibfnamefont{H.}~\bibnamefont{Sakaguchi}}
  \bibnamefont{et~al.}, \bibinfo{journal}{Phys. Rev. C}
  \textbf{\bibinfo{volume}{26}}, \bibinfo{pages}{944} (\bibinfo{year}{1982}).

\bibitem[{\citenamefont{Kato et~al.}(1985)}]{Ka85}
\bibinfo{author}{\bibfnamefont{S.}~\bibnamefont{Kato}} \bibnamefont{et~al.},
  \bibinfo{journal}{Phys. Rev. C} \textbf{\bibinfo{volume}{31}},
  \bibinfo{pages}{1616} (\bibinfo{year}{1985}).

\bibitem[{\citenamefont{Comfort et~al.}(1982)\citenamefont{Comfort, Moake,
  Foster, Schwandt, and Love}}]{Co82}
\bibinfo{author}{\bibfnamefont{J.~R.} \bibnamefont{Comfort}},
  \bibinfo{author}{\bibfnamefont{G.~L.} \bibnamefont{Moake}},
  \bibinfo{author}{\bibfnamefont{C.~C.} \bibnamefont{Foster}},
  \bibinfo{author}{\bibfnamefont{P.}~\bibnamefont{Schwandt}}, \bibnamefont{and}
  \bibinfo{author}{\bibfnamefont{W.~G.} \bibnamefont{Love}},
  \bibinfo{journal}{Phys. Rev. C} \textbf{\bibinfo{volume}{26}},
  \bibinfo{pages}{1800} (\bibinfo{year}{1982}).

\bibitem[{\citenamefont{Siefert et~al.}(1993)}]{Si93}
\bibinfo{author}{\bibfnamefont{H.}~\bibnamefont{Siefert}} \bibnamefont{et~al.},
  \bibinfo{journal}{Phys. Rev. C} \textbf{\bibinfo{volume}{47}},
  \bibinfo{pages}{1615} (\bibinfo{year}{1993}).

\bibitem[{\citenamefont{Hutcheon et~al.}(1981)}]{Hu81}
\bibinfo{author}{\bibfnamefont{D.~A.} \bibnamefont{Hutcheon}}
  \bibnamefont{et~al.}, in \emph{\bibinfo{booktitle}{AIP conf. Proc. No. 69}},
  edited by \bibinfo{editor}{\bibfnamefont{G.~G.} \bibnamefont{Ohlson}},
  \bibinfo{editor}{\bibfnamefont{R.~E.} \bibnamefont{Brown}},
  \bibinfo{editor}{\bibfnamefont{N.}~\bibnamefont{Jarmie}},
  \bibinfo{editor}{\bibfnamefont{W.~W.} \bibnamefont{McNaughton}},
  \bibnamefont{and} \bibinfo{editor}{\bibfnamefont{G.~M.} \bibnamefont{Hale}}
  (\bibinfo{publisher}{AIP}, \bibinfo{address}{New York},
  \bibinfo{year}{1981}).

\bibitem[{\citenamefont{Millburn et~al.}(1954)\citenamefont{Millburn, Birnbaum,
  Crandall, and Schecter}}]{Mi54}
\bibinfo{author}{\bibfnamefont{G.~P.} \bibnamefont{Millburn}},
  \bibinfo{author}{\bibfnamefont{W.}~\bibnamefont{Birnbaum}},
  \bibinfo{author}{\bibfnamefont{W.~E.} \bibnamefont{Crandall}},
  \bibnamefont{and} \bibinfo{author}{\bibfnamefont{L.}~\bibnamefont{Schecter}},
  \bibinfo{journal}{Phys. Rev.} \textbf{\bibinfo{volume}{95}},
  \bibinfo{pages}{1268} (\bibinfo{year}{1954}).

\bibitem[{\citenamefont{Cassels and Lawson}(1954)}]{Ca54}
\bibinfo{author}{\bibfnamefont{J.~M.} \bibnamefont{Cassels}} \bibnamefont{and}
  \bibinfo{author}{\bibfnamefont{J.~D.} \bibnamefont{Lawson}},
  \bibinfo{journal}{Proc. Phys. Soc. (London)} \textbf{\bibinfo{volume}{A67}},
  \bibinfo{pages}{125} (\bibinfo{year}{1954}).

\bibitem[{\citenamefont{Burge}(1959)}]{Bu59}
\bibinfo{author}{\bibfnamefont{E.~J.} \bibnamefont{Burge}},
  \bibinfo{journal}{Nucl. Phys.} \textbf{\bibinfo{volume}{13}},
  \bibinfo{pages}{511} (\bibinfo{year}{1959}).

\bibitem[{\citenamefont{Gooding}(1959)}]{Go59}
\bibinfo{author}{\bibfnamefont{T.~J.} \bibnamefont{Gooding}},
  \bibinfo{journal}{Nucl. Phys.} \textbf{\bibinfo{volume}{12}},
  \bibinfo{pages}{241} (\bibinfo{year}{1959}).

\bibitem[{\citenamefont{Meyer et~al.}(1960)\citenamefont{Meyer, Eisberg, and
  Carlson}}]{Me60}
\bibinfo{author}{\bibfnamefont{V.}~\bibnamefont{Meyer}},
  \bibinfo{author}{\bibfnamefont{R.~M.} \bibnamefont{Eisberg}},
  \bibnamefont{and} \bibinfo{author}{\bibfnamefont{R.~F.}
  \bibnamefont{Carlson}}, \bibinfo{journal}{Phys. Rev.}
  \textbf{\bibinfo{volume}{117}}, \bibinfo{pages}{1334} (\bibinfo{year}{1960}).

\bibitem[{\citenamefont{Johansson et~al.}(1961)\citenamefont{Johansson,
  Svanberg, and Sundberg}}]{Jo61}
\bibinfo{author}{\bibfnamefont{A.}~\bibnamefont{Johansson}},
  \bibinfo{author}{\bibfnamefont{U.}~\bibnamefont{Svanberg}}, \bibnamefont{and}
  \bibinfo{author}{\bibfnamefont{O.}~\bibnamefont{Sundberg}},
  \bibinfo{journal}{Arkiv Fysik} \textbf{\bibinfo{volume}{19}},
  \bibinfo{pages}{527} (\bibinfo{year}{1961}).

\bibitem[{\citenamefont{Goloskie and Strauch}(1962)}]{Go62}
\bibinfo{author}{\bibfnamefont{R.}~\bibnamefont{Goloskie}} \bibnamefont{and}
  \bibinfo{author}{\bibfnamefont{K.}~\bibnamefont{Strauch}},
  \bibinfo{journal}{Nucl. Phys.} \textbf{\bibinfo{volume}{29}},
  \bibinfo{pages}{474} (\bibinfo{year}{1962}).

\bibitem[{\citenamefont{Giles and Burge}(1954)}]{Gi64}
\bibinfo{author}{\bibfnamefont{R.~A.} \bibnamefont{Giles}} \bibnamefont{and}
  \bibinfo{author}{\bibfnamefont{E.~J.} \bibnamefont{Burge}},
  \bibinfo{journal}{Nucl. Phys.} \textbf{\bibinfo{volume}{50}},
  \bibinfo{pages}{327} (\bibinfo{year}{1954}).

\bibitem[{\citenamefont{Makino et~al.}(1964)\citenamefont{Makino, Waddell, and
  Eisberg}}]{Ma64}
\bibinfo{author}{\bibfnamefont{M.~Q.} \bibnamefont{Makino}},
  \bibinfo{author}{\bibfnamefont{C.~N.} \bibnamefont{Waddell}},
  \bibnamefont{and} \bibinfo{author}{\bibfnamefont{R.~M.}
  \bibnamefont{Eisberg}}, \bibinfo{journal}{Nucl. Phys.}
  \textbf{\bibinfo{volume}{50}}, \bibinfo{pages}{145} (\bibinfo{year}{1964}).

\bibitem[{\citenamefont{Kirkby and Link}(1966)}]{Ki66}
\bibinfo{author}{\bibfnamefont{P.}~\bibnamefont{Kirkby}} \bibnamefont{and}
  \bibinfo{author}{\bibfnamefont{W.~T.} \bibnamefont{Link}},
  \bibinfo{journal}{Can. J. Phys.} \textbf{\bibinfo{volume}{44}},
  \bibinfo{pages}{1847} (\bibinfo{year}{1966}).

\bibitem[{\citenamefont{Menet et~al.}(1971)\citenamefont{Menet, Gross,
  Malanify, and Zucker}}]{Me71}
\bibinfo{author}{\bibfnamefont{J.~J.~H.} \bibnamefont{Menet}},
  \bibinfo{author}{\bibfnamefont{E.~E.} \bibnamefont{Gross}},
  \bibinfo{author}{\bibfnamefont{J.~J.} \bibnamefont{Malanify}},
  \bibnamefont{and} \bibinfo{author}{\bibfnamefont{A.}~\bibnamefont{Zucker}},
  \bibinfo{journal}{Phys. Rev. C} \textbf{\bibinfo{volume}{4}},
  \bibinfo{pages}{1114} (\bibinfo{year}{1971}).

\bibitem[{\citenamefont{Renberg et~al.}(1972)\citenamefont{Renberg, Measday,
  Pepin, Schwaller, Favier, and Richard-Serre}}]{Re72}
\bibinfo{author}{\bibfnamefont{P.~U.} \bibnamefont{Renberg}},
  \bibinfo{author}{\bibfnamefont{D.~F.} \bibnamefont{Measday}},
  \bibinfo{author}{\bibfnamefont{M.}~\bibnamefont{Pepin}},
  \bibinfo{author}{\bibfnamefont{P.}~\bibnamefont{Schwaller}},
  \bibinfo{author}{\bibfnamefont{B.}~\bibnamefont{Favier}}, \bibnamefont{and}
  \bibinfo{author}{\bibfnamefont{C.}~\bibnamefont{Richard-Serre}},
  \bibinfo{journal}{Nucl. Phys.} \textbf{\bibinfo{volume}{A183}},
  \bibinfo{pages}{81} (\bibinfo{year}{1972}).

\bibitem[{\citenamefont{McGill et~al.}(1974)}]{Mc74}
\bibinfo{author}{\bibfnamefont{W.~F.} \bibnamefont{McGill}}
  \bibnamefont{et~al.}, \bibinfo{journal}{Phys. Rev. C}
  \textbf{\bibinfo{volume}{10}}, \bibinfo{pages}{2237} (\bibinfo{year}{1974}).

\bibitem[{\citenamefont{Slaus et~al.}(1975)\citenamefont{Slaus, Margaziotis,
  Carlson, van Oers, and Richardson}}]{Sl75}
\bibinfo{author}{\bibfnamefont{I.}~\bibnamefont{Slaus}},
  \bibinfo{author}{\bibfnamefont{D.~J.} \bibnamefont{Margaziotis}},
  \bibinfo{author}{\bibfnamefont{R.~F.} \bibnamefont{Carlson}},
  \bibinfo{author}{\bibfnamefont{W.~T.~H.} \bibnamefont{van Oers}},
  \bibnamefont{and} \bibinfo{author}{\bibfnamefont{J.~R.}
  \bibnamefont{Richardson}}, \bibinfo{journal}{Phys. Rev. C}
  \textbf{\bibinfo{volume}{12}}, \bibinfo{pages}{1093} (\bibinfo{year}{1975}).

\bibitem[{\citenamefont{Ingemarsson et~al.}(1999)}]{In99}
\bibinfo{author}{\bibfnamefont{A.}~\bibnamefont{Ingemarsson}}
  \bibnamefont{et~al.}, \bibinfo{journal}{Nucl. Phys.}
  \textbf{\bibinfo{volume}{A653}}, \bibinfo{pages}{341} (\bibinfo{year}{1999}).

\bibitem[{\citenamefont{Chapman and Macleod}(1967)}]{Ch67}
\bibinfo{author}{\bibfnamefont{R.}~\bibnamefont{Chapman}} \bibnamefont{and}
  \bibinfo{author}{\bibfnamefont{A.~M.} \bibnamefont{Macleod}},
  \bibinfo{journal}{Nucl. Phys.} \textbf{\bibinfo{volume}{A94}},
  \bibinfo{pages}{313} (\bibinfo{year}{1967}).

\bibitem[{\citenamefont{Carlson et~al.}(1975)}]{Ca75}
\bibinfo{author}{\bibfnamefont{R.~F.} \bibnamefont{Carlson}}
  \bibnamefont{et~al.}, \bibinfo{journal}{Phys. Rev. C}
  \textbf{\bibinfo{volume}{12}}, \bibinfo{pages}{1167} (\bibinfo{year}{1975}).

\bibitem[{\citenamefont{Turner et~al.}(1964)\citenamefont{Turner, Ridley,
  Cavanagh, Gard, and Hardacre}}]{Tu64}
\bibinfo{author}{\bibfnamefont{J.~F.} \bibnamefont{Turner}},
  \bibinfo{author}{\bibfnamefont{B.~W.} \bibnamefont{Ridley}},
  \bibinfo{author}{\bibfnamefont{P.~E.} \bibnamefont{Cavanagh}},
  \bibinfo{author}{\bibfnamefont{G.~A.} \bibnamefont{Gard}}, \bibnamefont{and}
  \bibinfo{author}{\bibfnamefont{A.~G.} \bibnamefont{Hardacre}},
  \bibinfo{journal}{Nucl. Phys.} \textbf{\bibinfo{volume}{58}},
  \bibinfo{pages}{509} (\bibinfo{year}{1964}).

\bibitem[{\citenamefont{Dicello and Igo}(1970)}]{Di70}
\bibinfo{author}{\bibfnamefont{J.~F.} \bibnamefont{Dicello}} \bibnamefont{and}
  \bibinfo{author}{\bibfnamefont{G.}~\bibnamefont{Igo}},
  \bibinfo{journal}{Phys. Rev. C} \textbf{\bibinfo{volume}{2}},
  \bibinfo{pages}{488} (\bibinfo{year}{1970}).

\bibitem[{\citenamefont{Wilkins and Igo}(1963)}]{Wi63}
\bibinfo{author}{\bibfnamefont{B.~D.} \bibnamefont{Wilkins}} \bibnamefont{and}
  \bibinfo{author}{\bibfnamefont{G.}~\bibnamefont{Igo}},
  \bibinfo{journal}{Phys. Rev.} \textbf{\bibinfo{volume}{129}},
  \bibinfo{pages}{2198} (\bibinfo{year}{1963}).

\bibitem[{\citenamefont{Dicello et~al.}(1967)\citenamefont{Dicello, Igo, and
  Roush}}]{Di67}
\bibinfo{author}{\bibfnamefont{J.~F.} \bibnamefont{Dicello}},
  \bibinfo{author}{\bibfnamefont{G.~J.} \bibnamefont{Igo}}, \bibnamefont{and}
  \bibinfo{author}{\bibfnamefont{M.~L.} \bibnamefont{Roush}},
  \bibinfo{journal}{Phys. Rev.} \textbf{\bibinfo{volume}{157}},
  \bibinfo{pages}{1001} (\bibinfo{year}{1967}).

\bibitem[{\citenamefont{Pollock and Schrank}(1965)}]{Po65}
\bibinfo{author}{\bibfnamefont{R.~E.} \bibnamefont{Pollock}} \bibnamefont{and}
  \bibinfo{author}{\bibfnamefont{G.}~\bibnamefont{Schrank}},
  \bibinfo{journal}{Phys. Rev.} \textbf{\bibinfo{volume}{140}},
  \bibinfo{pages}{B575} (\bibinfo{year}{1965}).

\bibitem[{\citenamefont{Montague et~al.}(1973)\citenamefont{Montague, Cole,
  Makino, and Waddell}}]{Mo73}
\bibinfo{author}{\bibfnamefont{D.~G.} \bibnamefont{Montague}},
  \bibinfo{author}{\bibfnamefont{R.~K.} \bibnamefont{Cole}},
  \bibinfo{author}{\bibfnamefont{M.}~\bibnamefont{Makino}}, \bibnamefont{and}
  \bibinfo{author}{\bibfnamefont{C.~N.} \bibnamefont{Waddell}},
  \bibinfo{journal}{Nucl. Phys.} \textbf{\bibinfo{volume}{A199}},
  \bibinfo{pages}{457} (\bibinfo{year}{1973}).

\bibitem[{\citenamefont{Amos et~al.}(2003)\citenamefont{Amos, Canton, Pisent,
  Svenne, and van~der Knijff}}]{Am03}
\bibinfo{author}{\bibfnamefont{K.}~\bibnamefont{Amos}},
  \bibinfo{author}{\bibfnamefont{L.}~\bibnamefont{Canton}},
  \bibinfo{author}{\bibfnamefont{G.}~\bibnamefont{Pisent}},
  \bibinfo{author}{\bibfnamefont{J.~P.} \bibnamefont{Svenne}},
  \bibnamefont{and} \bibinfo{author}{\bibfnamefont{D.}~\bibnamefont{van~der
  Knijff}}, \bibinfo{journal}{Nucl.\ Phys.} \textbf{\bibinfo{volume}{A728}},
  \bibinfo{pages}{65} (\bibinfo{year}{2003}).

\bibitem[{\citenamefont{Abfalterer et~al.}(2001)\citenamefont{Abfalterer,
  Bateman, Dietrich, Finaly, Haight, and Morgan}}]{Ab01}
\bibinfo{author}{\bibfnamefont{W.~P.} \bibnamefont{Abfalterer}},
  \bibinfo{author}{\bibfnamefont{F.~B.} \bibnamefont{Bateman}},
  \bibinfo{author}{\bibfnamefont{F.~S.} \bibnamefont{Dietrich}},
  \bibinfo{author}{\bibfnamefont{R.~W.} \bibnamefont{Finaly}},
  \bibinfo{author}{\bibfnamefont{R.~C.} \bibnamefont{Haight}},
  \bibnamefont{and} \bibinfo{author}{\bibfnamefont{G.~L.}
  \bibnamefont{Morgan}}, \bibinfo{journal}{Phys. Rev. C}
  \textbf{\bibinfo{volume}{63}}, \bibinfo{pages}{044608}
  (\bibinfo{year}{2001}).

\bibitem[{\citenamefont{Finlay et~al.}(1993)\citenamefont{Finlay, Abfalterer,
  Fink, Montei, Adami, Lisowski, Morgan, and Haight}}]{Fi93}
\bibinfo{author}{\bibfnamefont{R.~W.} \bibnamefont{Finlay}},
  \bibinfo{author}{\bibfnamefont{W.~P.} \bibnamefont{Abfalterer}},
  \bibinfo{author}{\bibfnamefont{G.}~\bibnamefont{Fink}},
  \bibinfo{author}{\bibfnamefont{E.}~\bibnamefont{Montei}},
  \bibinfo{author}{\bibfnamefont{T.}~\bibnamefont{Adami}},
  \bibinfo{author}{\bibfnamefont{P.~W.} \bibnamefont{Lisowski}},
  \bibinfo{author}{\bibfnamefont{G.~L.} \bibnamefont{Morgan}},
  \bibnamefont{and} \bibinfo{author}{\bibfnamefont{R.~C.}
  \bibnamefont{Haight}}, \bibinfo{journal}{Phys. Rev. C}
  \textbf{\bibinfo{volume}{47}}, \bibinfo{pages}{237} (\bibinfo{year}{1993}).

\bibitem[{\citenamefont{Amos et~al.}(2002)\citenamefont{Amos, Karataglidis, and
  Deb}}]{Am02}
\bibinfo{author}{\bibfnamefont{K.}~\bibnamefont{Amos}},
  \bibinfo{author}{\bibfnamefont{S.}~\bibnamefont{Karataglidis}},
  \bibnamefont{and} \bibinfo{author}{\bibfnamefont{P.~K.} \bibnamefont{Deb}},
  \bibinfo{journal}{Phys. Rev. C} \textbf{\bibinfo{volume}{65}},
  \bibinfo{pages}{064618} (\bibinfo{year}{2002}).

\bibitem[{\citenamefont{Deb et~al.}(2004)\citenamefont{Deb, Amos, and
  Karataglidis}}]{De04}
\bibinfo{author}{\bibfnamefont{P.~K.} \bibnamefont{Deb}},
  \bibinfo{author}{\bibfnamefont{K.}~\bibnamefont{Amos}}, \bibnamefont{and}
  \bibinfo{author}{\bibfnamefont{S.}~\bibnamefont{Karataglidis}},
  \bibinfo{journal}{Phys. Rev. C} \textbf{\bibinfo{volume}{70}},
  \bibinfo{pages}{057601} (\bibinfo{year}{2004}).

\bibitem[{\citenamefont{Cohen et~al.}(1991)\citenamefont{Cohen, Furnstahl, and
  Griegel}}]{Co91}
\bibinfo{author}{\bibfnamefont{T.~D.} \bibnamefont{Cohen}},
  \bibinfo{author}{\bibfnamefont{R.~J.} \bibnamefont{Furnstahl}},
  \bibnamefont{and} \bibinfo{author}{\bibfnamefont{D.~K.}
  \bibnamefont{Griegel}}, \bibinfo{journal}{Phys. Rev. Lett.}
  \textbf{\bibinfo{volume}{67}}, \bibinfo{pages}{961} (\bibinfo{year}{1991}).

\end{thebibliography}

\end{document}